
%

\magnification 1200

\def\IN{\relax{\rm I\kern-.18em N}}
\def\onequal#1{\mathrel{\mathop{=}\limits^{#1}}}
\def\ll{\left\langle}
\def\rr{\right\rangle}
\def\svec#1{\skew{-2}\vec#1}
\def\undermax#1{\mathrel{\mathop{\max}\limits_{#1}}}
\def\onarrow#1{\mathrel{\mathop{\longrightarrow}\limits^{#1}}}
\def\lessim%
{\lower0.6ex\hbox{\vbox{\offinterlineskip\hbox{$<$}\vskip1pt\hbox{$\sim$}}}}
\def\grtsim%
{\lower0.6ex\hbox{\vbox{\offinterlineskip\hbox{$>$}\vskip1pt\hbox{$\sim$}}}}

\overfullrule 0pt

\def\footnoterule{\kern-3pt \hrule width \hsize \kern2.6pt}
\pageno=0
\footline={\ifnum\pageno>0 \hss --\folio-- \hss \else\hfil\fi}

\baselineskip 14pt plus 1pt minus 1pt

\centerline{\bf CALCULATION AND INTERPRETATION OF HADRON}
\centerline{{\bf CORRELATION FUNCTIONS IN LATTICE QCD}\footnote{*}
{This work is supported in part by funds provided by the U. S.
Department of Energy (D.O.E.) under contracts DE-FG06-90ER40561,
DE-FG06-88ER40427, and DE-FC02-94ER40818.}}
\vskip 24pt
\centerline{M.~Burkardt\footnote{$\ddagger$}
{Present address:
Institute for Nuclear Theory,
University of Washington,
Seattle WA ~98195},
J. M. Grandy\footnote{$\dagger$}
{Present address:
MS B-285,
Los Alamos National Laboratory,
Los Alamos NM ~87545}
and J.~W.~Negele}
\vskip 12pt
\centerline{\it Center for Theoretical Physics}
\centerline{\it Laboratory for Nuclear Science}
\centerline{\it and Department of Physics}
\centerline{\it Massachusetts Institute of Technology}
\centerline{\it Cambridge, Massachusetts\ \ 02139\ \ \ U.S.A.}
\vfill
\centerline{Submitted to: {\it Annals of Physics}}
\vfill
\line{CTP\#2109 \hfil hep-lat/9406009 \hfil May 1994}
\eject

\baselineskip 20pt plus 2pt minus 2pt
\centerline{\bf ABSTRACT}
\medskip

Several new developments in the calculation and interpretation of
hadron density-density correlation functions are presented.  The
asymptotic behavior of correlation functions is determined from a tree
diagram path integral. A method is developed to use this behavior to
correct for leading image contributions on a finite periodic spatial
lattice and to correct for the finite temporal extent of the lattice.
Equal time correlation functions are shown to determine a sum of the
ground state rms radius plus a polarization correction, and it is
shown how to extract the hadron polarizability from unequal time
correlation functions.  Image-corrected correlation functions
calculated in quenched lattice QCD are presented and used to determine
the size of the pion and nucleon.

\vfill\eject
\noindent{\bf I.\quad INTRODUCTION}\medskip\nobreak
One of the primary motivations for solving QCD on a lattice is to understand
the structure of hadrons. Hence, in addition to reproducing the hadron
spectrum and experimentally measured form factors, it is important to explore
the spatial distributions of quarks and the correlations between them in the
ground states of hadrons. As in other strongly interacting many-body systems,
two-body correlation functions are a natural starting point, with the simplest
being the density-density correlation function
$$
\rho({\vec y}, t_1, t_2) \equiv \int d{\vec x} \left<h|
\bar{u}_{{\vec x},t_1}\gamma^0 u_{{\vec x},t_1}
\bar{d}_{{\vec x}+{\vec y},t_2}\gamma^0 d_{{\vec x}+{\vec y},t_2}
|h\right>
\eqno(1.1)
$$
where $|h\rangle$
denotes a hadronic ground state and $u$ and $d$ represent field
operators for up and down quarks respectively.  Although these correlation
functions have been studied in a number of works [1--4], several important
practical and conceptual problems were previously unresolved and
errors were made in interpreting correlation functions. Hence, this work
presents a number of new results relevant to the calculation and
interpretation of density-density correlation functions.

One practical problem surrounding all lattice QCD calculations is the
irreconcilable conflict with finite computer resources between reducing the
lattice spacing (and thus accurately approximating the continuum limit) and
increasing the physical volume of the lattice (and thus eliminating finite
volume corrections). Nowhere is the conflict more severe than in the
calculation of density-density correlation functions in hadrons.  Conventional
hadron wave functions, either gauge fixed in Coulomb or Landau gauge or made
gauge invariant by a line of flux $e^{i\int dx \, A}$,
fall off much more rapidly
than quark correlation functions because they contain additional suppression
factors at large separation arising from the small overlap between the gluon
wave functional in the hadron and in the vacuum [5]. Typically the
spatial extent of these wave functions is half that of the density-density
correlation functions, so that lattices in common use which are adequate for
wave functions are much too small for correlation functions [3].

In section 2, we therefore take a fresh look at the general problem of
relating correlation functions in a periodically replicated array of
hadrons, corresponding to periodic boundary conditions on a spatial
lattice, to those in an isolated hadron. We begin with a pedagogical
example in one spatial dimension and show how to correct for the
effects of periodic images and generalize to the physical case of
three space dimensions.  A crucial ingredient in obtaining the
properties of an isolated hadron is understanding the asymptotic decay
of the correlation functions. Since the behavior is more complicated
than simple exponential decay governed by the rho mass, in section 3
we construct a tree diagram path integral which incorporates the
essential physics.  This path integral accurately describes the
asymptotic behavior and provides a means to remove finite spatial and
temporal volume effects. It also explains an apparent paradox wherein
two seemingly equivalent methods of calculating correlation functions
yielded different asymptotic behavior [6].

Given the possibility of accurately determining the density-density
correlation function for an isolated hadron, it is important to understand
clearly its physical content and to explore the full range of physical
information which can be extracted from it. For this reason, in section IV we
reexamine the physical content of the density-density correlation functions at
equal and unequal time in the nonrelativistic limit, which is particularly
simple to understand.  We show that the rms radius of the equal time
correlation function is not the rms radius of the hadron, as previously
claimed, but rather is the sum of the rms radius and a dipole-dipole
polarization term.  Furthermore, we show how to extract the hadron
polarizability from integrating the correlation function over relative time,
and generalize the results to the relativistic case and more general
currents.

Finally, having developed the general formalism and methodology,
the results of quenched lattice calculations for density-density correlation
functions for the $\pi$, $\rho$ and nucleon are presented and discussed
in section~V.
The conclusions   and outlook   are discussed in section VI.

\vfill\eject

\noindent{\bf II.\quad PERIODIC IMAGES IN LATTICE GAUGE THEORIES}
\medskip\nobreak

Intuitively, it is clear that when we calculate hadron correlation functions
in a finite spatial volume with periodic boundary conditions, we are not
considering an isolated hadron but rather a periodically replicated array of
hadrons.  Consequently, periodicity effects and interactions between
neighboring hadrons will occur and we either must discover how to correct for
these effects or calculate with a volume so large that these effects are
negligible.

The primary result we will establish in this section is that the
density-density correlation function for a gauge theory on a periodic lattice
of spatial dimension $L$ has the form
$$
\rho_{\rm periodic} (\vec{r}) = \sum_{\vec{n}} \rho_0 (\vec{r}+\vec{n}L)
\eqno{(2.1)}
$$
where $\rho_0(\vec{r})$ is an accurate approximation to the free hadron result
$\rho(\vec{r})$ for $r < L$ and differs due to interactions with periodic
images only for $r \grtsim L$.  This result embodies two physically distinct
effects.  The dominant effect of periodicity is just the
summation of the tails of $\rho_0$ from neighboring images, and this effect
amounts to a factor of 2 correction at distance ${L \over 2}$ along the
Cartesian axes.
The second effect is the discrepancy between $\rho_0$ and $\rho$ arising from
interactions with periodic images, and it first appears at distance $L$.
Fortunately, since the dominant effect is the sum of tails from images, it can
be removed by subtracting these image contributions, leaving only the residual
discrepancy between $\rho_0$ and $\rho$ beyond $r \approx L$ which will be
negligible in cases of practical interest.

 To understand how the form (2.1) arises in a gauge theory and how
$\rho_0$ may be obtained by subtracting image contributions, it is
useful to begin by considering the simple example of QED coupled to
non-relativistic particles in $1+1$ dimensions on a circle.

\bigskip
\noindent {\bf Positronium on a circle} \smallskip\nobreak

In the continuum, using the gauge $\partial_x A_x=0$, the Hamiltonian
for a non-relativistic
$e^+e^-$ pair in $1+1$ dimensions on a circle with circumference $L$ is
$$H = - {1\over 2L}\ {d^2\over dA^2_x} + {\left( p_1 - e A_x\right)^2\over 2m}
+ {\left( p_2 + eA_x\right)^2\over 2m} + V\left(
x_1-x_2\right)\eqno(\hbox{2.2})$$
where $V(x)$ is the periodic potential
$$\eqalign{V(x) &= {e^2\over 2\pi}\ {L\over 2\pi}
\sum^\infty_{{n=-\infty}\atop{n\not=0}} {1-e^{\displaystyle{2\pi i xn\over
L}}\over n^2}\cr
 &= {e^2 \over 2} |\tilde{x}|
\left( 1 - {|\tilde{x}|\over L}\right)\ \ ,\qquad
\tilde{x} = |x|_{mod \, L} \cr}\eqno(\hbox{2.3})$$
The Schr\"odinger equation corresponding to $H$, Eq.~(2.2), must be solved with
appropriate boundary conditions on the wavefunction. Here we will use
periodic boundary conditions  in $x_1$ and $x_2$
$$\psi\left(x_1+L, x_2,A_x\right) = \psi \left( x_1, x_2+L,A_x\right) = \psi
\left( x_1, x_2, A_x\right) \eqno(\hbox{2.4})$$
although other boundary conditions are also possible.  The
center-of-mass motion separates in Eq.~(2.2), {\it i.e.\/}
$$H = - {1\over 2L}\ {d^2\over dA^2_x} + {P^2\over 4m} + {\left( p - e
A_x\right)^2\over m} + V(x) \eqno(\hbox{2.5})$$
where $P = p_1 + p_2$, $p= (p_1-p_2)/2$, $x=x_1-x_2$.  Notice that Eq.~(2.4)
implies periodicity in the center-of-mass and the relative coordinate
separately.  The eigenstates of $H$ corresponding to $P=0$ can be
expressed in the form
$$\psi_k (x,A_x) = \sum_n e^{i(x-nL) A_x\cdot e} \hat\psi_k (x-nL)\ \ .
\eqno(\hbox{2.6})$$
where $\hat\psi_k$ is a solution of the equation
$$E_k \hat\psi_k (x) = \left( {p^2\over m} + \hat V(x)\right) \hat\psi_k(x)\ \
,\eqno(\hbox{2.7})$$
with
$$\hat V(x) = V(x) + {e^2\over 2L} x^2\ \ .\eqno(\hbox{2.8})$$
Note that $\psi (x,A_x)$ is defined only for $-L<x<L$ (with periodic
or quasiperiodic boundary conditions), whereas ${\hat \psi}(x)$ is defined
for $-\infty < x < \infty$ (with vanishing boundary conditions).  This
difference in domain allows a function of one variable to carry the
same information as a function of two variables.

For $x = nL + \Delta x$, where $n\in\IN$ and $0<\Delta x<L$, the effective
potential $\hat V$ can be rewritten in the form
$$\hat V (x) = {e^2\over 2L} (nL)^2 + (2n+1) {e^2\over 2}\Delta x\ \
,\eqno(\hbox{2.9})$$
which allows a simple physical interpretation.  Consider the case in
which the $e^+$ moves around the circle $n$ times while the $e^-$ is
fixed.  The electric flux from the $e^+$ to the $e^-$ ends up in a
configuration where it is wrapped around the circle $n$ times, since
there are no terms in the Hamiltonian which would allow the gauge
field to make a  transition to the energetically lowest unwrapped
configuration.  The electric field energy density is proportional to
the electric field strength squared --- the latter being $(n+1)\cdot
e$ for $0<x<\Delta x$ and $ne$ for $\Delta x<x<L$ in the above
example.  Hence
$$E_{\rm gauge} = {1\over 2} \left\{ \Delta x\cdot \left[ (n+1)e\right]^2 +
(L-\Delta x) [ne]^2\right\} \eqno(\hbox{2.10})$$
which agrees with Eq.~(2.9).

Note that for $-L<x<L$ the effective potential (which determines the
excitation spectrum of $H$) agrees with the potential on a line, {\it i.e.\/}
$\hat V(x) = {e^2\over 2}|x|$. In contrast, if we had  omitted the
$A_x$ degree of freedom, we would have obtained ($|x|<L$)
$$V(x) = {e^2\over 2} |x| \left( 1 - {|x|\over L}\right)\ \ ,\qquad
\hbox{(periodic)}\eqno(\hbox{2.11})$$
which is a much worse approximation to ${e^2\over 2}|x|$ than Eq.~(2.9).
Although the $A_x$ degree of freedom ``freezes out'' for $L\to \infty$ (see
Eq.~(2.5)), its presence improves the approach to the infinite volume
result for finite $L$.

For $|x|>L$, the interaction is modified by the periodic boundary
condition.  In a sense the positronium interacts with its own periodic
replicas.  However, in most cases, this has only an exponentially
suppressed effect on energy eigenvalues and the wavefunctions, since
the wave functions fall off exponentially for large $x$.

For ground state correlations another finite size effect is of much more
importance.  Consider
$$\rho_k(x) \equiv
\int dy \ll k| \delta(x_1-y) \delta(x_2-x-y)|k\rr\ \ ,\eqno(\hbox{2.12})$$
where $|k\rangle$ denotes the $k^{\rm th}$ positronium state.  Using Eq.~(2.6),
$$\eqalign{\rho_k(x) &\propto\int
 dA\left| \psi_k (A,x)\right|^2 = \sum_{n,m} \int
dA\, e^{i(mL-nL) eA}\hat\psi_k (x-nL) \hat\psi^*_k(x-mL) \cr
&= 2\pi \sum_{n,m} \delta(nL - mL) \hat\psi_k (x-nL) \hat\psi^*_k(x-mL) \cr
&\propto \sum_n \hat\rho_k (x-nL)\ \ ,\cr}\eqno(\hbox{2.13})$$
where
$$\hat\rho_k(x) = \left| \hat\psi_k (x)\right|^2 \ \ .\eqno(\hbox{2.14})$$
The essential result is that the interference terms ($\propto
\hat\psi_n\hat\psi^*_m$, $n\not=m$) are removed by the integration
over $A$.\footnote{*}{To avoid irrelevant details, we have not been
explicit about normalization and the $\delta(0)$-term ($m=n$ in
Eq.~(2.13)) arising because we work in the continuum. In a lattice
calculation, the gauge field integration is compact.}

Equation (2.13) shows explicitly how the general result, Eq.~(2.1), arises
in $1+1$ dimensions, gives a precise definition of $\rho_0$ as
$\hat\rho_k$, and displays the role the gauge field plays in
obtaining this result. The longitudinal gauge field $A_x$, which is
the only quantum mechanical degree of freedom associated with the
gauge field in QED$_{1+1}$, plays two essential roles.  First, it
removes all cross terms so only the square of the wave function
$|\tilde\psi_k(x)|^2$ appears in the final result.  Secondly, it
replaces the periodic potential, Eq.~(2.11), by the linear potential
in the range $-L < x < L$, so that $\tilde\psi_k(x)$ satisfies the
correct Schr\"odinger equation in this entire range.  The role of
$A_x$ has also been emphasized by a number of authors in the context
of the Schwinger model [7], where it turns out to be crucial in
providing a proper description of the anomaly in the divergence of the
axial vector current.

\goodbreak\bigskip
\noindent {\bf A numerical example  in 1+1 dimensions}
\smallskip\nobreak
To show how the density for a free bound state, $\rho_0(x)$, may be
extracted from the periodic sum $\rho_{\rm periodic}(x)$ in Eq.~(2.1), we
now consider a numerical example.  For convenience, and to establish
contact with the lattice formulation of QCD$_{3+1}$, we use a discrete
lowest order difference approximation to the Hamiltonian, Eq.~(2.7).
Note that beginning with QED$_{1+1}$ lattice gauge theory and gauge
fixing analogously yields an equivalent form with one quantum
mechanical gauge degree of freedom remaining [8,9].
The general conclusion will be independent of the discretization.

To emphasize the image corrections, we have chosen the example shown in
Fig.~1, in which there is greater overlap between the bound state and
its periodic images than in a typical lattice QCD calculation.  Note
that the periodic solution , denoted by the heavy solid curve, falls
to less than half of its peak value.  This case has $N=10$ lattice sites
separated by spacing $a$, with mass $ma=0.3$ and charge $ea=0.3$.
With these parameters, the exact solution on the open line, shown by
the light solid curve, has size $\sqrt{\ll r^2\rr}\simeq\hbox{3.6a}$.

Our general strategy for image corrections is to use the known
asymptotic decay of the infinite domain solution both to subtract the tails
of the first images in the fundamental unit cell and to approximate
the solution outside the unit cell. Thus, in any dimension, assuming
the asymptotic form $\rho_{\rm as}({\vec r})$ is known, the image
corrected density $\rho_{\rm cor}({\vec r})$ is defined
$$\rho_{\rm cor} ({\svec r}) = \rho_{\rm as}({\svec r})\ \ \eqno(2.15a)$$
for ${\svec r}$ outside the fundamental unit cell and
$$\rho_{\rm cor} ({\svec r}) = \rho_{\rm periodic}({\svec
r})-\sum_{{\svec n}\neq0}
\rho_{\rm as}({\svec r + \svec n L})\ \ \eqno(\hbox{2.15b}) $$
for ${\svec r}$ inside the first unit cell.

In our one-dimensional example,  we make use of the known asymptotic behavior
for Airy functions, so that as  $a\to0$, the density has the form
$$\rho_{\rm as}(x) \sim \exp \left( - \lambda x^{3/2}\right)\ \
,\qquad\hbox{where}\qquad
\lambda = {4\over 3}\  \sqrt{{me^2\over 2}}\ \ .\eqno(\hbox{2.16})$$
In the continuum limit, with the above parameters, $\lambda = 0.155$.
For finite lattice spacing, $\lambda$ is somewhat smaller. The
normalization of $\rho_{\rm as}(x)$ is determined by the symmetry condition
$$\rho_{\rm as}({L\over2}) = {1\over2} \, \rho_{\rm
periodic}({L\over2})\ \ \eqno(\hbox{2.17})$$
Although, by construction, $\rho_{\rm cor}(x)$ and its first
derivative are continuous at $x = \pm
{L\over 2}$, its second derivative is not, both because of finite $a$
corrections to $\lambda$ and because $\rho(x)$ is not yet asymptotic
at $x = \pm {L\over 2}$. Especially when one is also interested in the
Fourier transform, it is desirable to parameterize the image
corrections such that the second derivative is continuous, and there
are several possibilities. One
possibility
is simply to define  $\lambda$  such that $\rho_{\rm cor}(x)$
also has a continuous second
derivative. Alternatively, one
may make the {\it ansatz\/}
$$
\rho_{\rm as}(x) \sim (a+bx) \exp\left( - \lambda x^{3/2}\right)
\eqno(2.18)
$$
and determine $b/a$ by the same criterion. Since there are polynomial
corrections to Eq.~(2.16), this form is clearly reasonable. The results
are essentially the same, and we only show calculations using the {\it
ansatz\/} (2.18).

The result of correcting the periodic solution using Eq.~(2.15) with
the asymptotic form Eq.~(2.18) with $\lambda= 0.13$ is shown by the
dashed line in Fig. 1.  This corrected result agrees quite well with
the exact solution on the infinite domain, shown by the light solid
line, especially considering the high degree of overlap in our
deliberately chosen worst-case example. Note that the reconstruction
of the density is not extremely sensitive to the precise value of
$\lambda$. Although the discrete solution on the open line yields
$\lambda\approx 0.145$, the value $\lambda=0.13$ we have chosen and
$\lambda=0.15$ used in calculations which are not shown here yield very
similar results.

The Fourier transforms of the densities in Fig.~1 are shown in Fig.~2.
We will show in Section IV that they are the sums of squares of elastic
and inelastic form factors. The Fourier transform of the periodic
density is only defined at the discrete lattice momenta $q_k = k\pi/L$
and is indicated by the solid dots in Fig.~2.
An equivalent way of
rephrasing our image correction problem is to ask whether we can use
our knowledge of the form of the density, Eq.~(2.1) and the asymptotic
behavior to reconstruct the Fourier transform for all $qa$. Clearly,
the curvature at the origin, for example, is related to the rms radius
of the correlation function and must depend sensitively on the details
of the surface in a way which cannot be guessed by naive polynomial
interpolation of the solid points. One observes that the Fourier
transform of the corrected density from Fig.~1, denoted by the solid
curve in part (a), agrees well with the Fourier transform of the exact
result on the open line. We should emphasize that the level of
agreement shown in part (a) of Fig.~2 and in Fig.~1 is not unique to
the particular set of parameters we have chosen, and that similar
results were obtained for a wide range of parameters $N$, $ma$, and $ea$.

The importance of making the second derivative of the corrected
density continuous is shown by
comparing curves (a) and (b) of Fig.~2.
The solid curve (b)
shows the result of using Eq.~(2.16) for the asymptotic tail with the
same value $\lambda=0.13$ but not introducing an additional linear term
as in Eq.~(2.18) which can be used to make the second derivative
continuous. The substantial oscillations around the exact result show
how large the effects from the discontinuity become at intermediate
values of $qa$. One should note that Eq.~(2.1) has the property that at
the discrete lattice momenta $q_k = k\pi/L$, the Fourier transforms of
$\rho_{\rm periodic}$ and $\rho_0$ are identical, so that to the
extent that second and higher image corrections are negligible, both
the solid and dashed lines must go through the solid dots.

In practical lattice calculations, the periodic signal
$\rho_{k}(x)$ in Eq.~(2.13)
is not a smooth function but rather is contaminated
with statistical noise.  This adds another level of complication to
the reconstruction of $\rho_{\rm cor}(x)$. To study this case, in part
(c) of Fig. 2 we have considered an ensemble of 10 data sets obtained
by adding randomly distributed errors with mean fractional deviation
1\% to the periodic density at each lattice site.  As before, for the
matching conditions we use continuity of $\rho$ and the second
derivative. Instead of using a three-point formula, both are now
computed using a five-point formula which automatically provides some
averaging over the statistical fluctuations. One observes that for
$q<\pi/L$ the resulting form factor turns out to be quite stable, and
one has as before an accurate determination of the rms radius.
However, as expected, for larger values of $q$ at which the Fourier
transform has fallen by two orders of magnitude, the 1\% errors
introduce uncertainties comparable to the signal. In this region, the
uncertainty is well represented by the variance of the solid dots
coming from the periodic solution and no further information can be
obtained by image corrections.

\goodbreak\medskip
\noindent {\bf Image Corrections in ${\rm QCD}_{3+1}$}
\smallskip\nobreak
Fortunately, although one cannot carry through an explicit solution as in the
case of ${\rm QED}_{1+1}$, Eq.~(2.1) also applies to ${\rm QCD}_{3+1}$ with
periodic boundary conditions in the spatial direction.
The effects of images may be
understood by considering all possible contractions of the field operators
occurring in all periodic replicas of the density operators  and all periodic
replicas of the sources.

Representative contractions relevant to meson density-density correlation
functions are sketched in Fig.~3, where one periodic spatial dimension $x$ is
shown on a cylinder and the other two spatial dimensions are suppressed.
The circumference of the cylinder is $L$, and the separation between the two
density operators on the midplane of the cylinder measured on the front of the
cylinder is denoted $r$ and around the back of the cylinder is $|r-L|$.
The upper sketch shows the case in which all contractions occur within the
fundamental unit cell and yields the physical result in the infinite volume
limit.   The middle sketch shows a typical contraction in which the meson is
created in the fundamental unit cell on the left creating the wave function
$\psi_0(r)$ at the midplane whereas on the right, a propagator from the first
periodic image creates the wave function $\psi_0(r-L)$ at the mid plane.  For
a periodic wave function which could be written in the form
$\psi(\vec{r}) = \sum_n \psi_0 (\vec{r} + \vec{n} L)$,
this contraction would thus correspond to a cross term of the form
$\psi_0 (\vec{r} + \vec{n} L) \, \psi_0(\vec{r} + \vec{m} L)$.  Clearly, if
this term were non-vanishing, one would never obtain the form Eq.~(2.1) which
is just the sum of diagonal terms.  However, since the total propagator in the
middle sketch has the topology of a spatial Polyakov line, it therefore
vanishes in the confining phase for a sufficiently large lattice.         This
is the physical reason no cross terms occur in gauge theories in any
dimension.   Finally, the bottom sketch shows a typical contraction in which
the meson is created with propagators from the first periodic image on the
left and on the right, corresponding to the diagonal term
$\psi_0(r-L) \, \psi_0(r-L)$,
giving rise to a diagonal contribution to the density of the form
$\rho(r+nL)$.
Extending this argument to all contractions in 3+1 dimensions, one can see
that the Polyakov line argument removes all cross terms, and that the result
must have the form of Eq.~(2.1).

Thus, in principle, the procedure for correcting images in higher dimensions
is completely analogous to that in two dimensions.  However, from the
practical point of view there is an additional complication since the boundary
of a cube in $D>1$ dimensions is a $D-1$-dimensional extended object and it
would be very difficult to impose continuity everywhere on the boundary.
Therefore, it is useful to consider more general matching conditions to
replace the continuity requirement.

To this end, it is important to note that there are only a few
critical points on the boundary where it is
essential
to achieve
continuity.
This can be seen by calculating
the angular averaged form factor
$$\tilde\rho(q) = \int d\Omega_q \tilde\rho ({\svec q}) \propto \int^\infty_0
dr\,r^{D-1} {\sin qr\over qr} \rho(r) ~~,
\eqno(2.19)
$$
where
$$
\rho(r) = \int d\Omega_r \rho({\svec r}) ~~.
\eqno(2.20)
$$
Obviously, if one wants to avoid unphysical oscillations in $\tilde\rho(q)$,
one should keep $\rho(r)$ as smooth as possible.  In order to see what
conditions this implies for $\rho({\svec r})$,
consider the extreme case of a step
function in two dimensions (${\svec r} = (x,y)$)
$$
\rho({\svec r}) = \cases{ 0 & $|x|>a$ or $|y|>a$\cr
1 & $|x|<a$ and $|y|<a$ \cr}
\eqno(2.21)
$$
yielding, after taking the angular average,
$$
\rho(r) = \cases{1 & $0\le r\le a$ \cr
1 - {\displaystyle{4\over\pi}} \sin^{-1} {\displaystyle{r\over 2a}}
& $a<r<a\cdot\sqrt{2}$\cr
0 & $r\ge a\cdot \sqrt{2}$\cr}
\eqno(2.22)
$$
Note that although $\rho({\svec r})$ is
discontinuous everywhere on the boundary of the square, $\rho(r)$ and its
derivatives are continuous
everywhere
except at two critical points: $r=a$ and
$r=\sqrt{2}\,a$, where $\rho'(r)$ is discontinuous.
Physically, this corresponds to that fact that when one increases $r$
and the sphere $|{\svec r}|=r$ intersects with the corner or the center of the
faces of the square (Eq.~(2.22)) then there is a
sudden change in the
fraction of the sphere contained inside the square, reflected in a
discontinuous  derivative.   Turning now   to three dimensions,   we see that
if we need  to match solutions inside and outside the cubic unit cell where
discontinuities can occur only at the boundary,
discontinuities in $\rho(r)$ or its
derivative will arise only
if $\rho({\svec r})$ is not continuous in the corners,
the center of the edges, or the center of the faces
of the first unit cell.  Hence, unphysical oscillations in
the form factor at large $q$ are avoided most efficiently if
the parameters of the asymptotic density, such as in Eq.~(2.18),
are determined such that the discontinuities around these
critical points are minimized.

In the practical case of lattice calculations in which the density measured at
every point has statistical errors, a different strategy is required, and an
iterative self-consistent method to determine the optimal density is described
in section V.

\goodbreak\bigskip
\noindent
{\bf III.\quad THE ASYMPTOTIC BEHAVIOR OF DENSITY-DENSITY CORRELATIONS}
\medskip\nobreak
In the last section we showed how one can use the knowledge of the
asymptotic behavior of the density for an isolated hadron,
$\rho_0({\svec r})$, to correct for
images in
the density of a periodic system,
related to $\rho_0$ by
$$\rho_{\rm periodic}({\svec r}) =
\sum\limits_{\svec n} \rho_0 \left( {\svec r} +
{\svec n} L\right) \ \ .\eqno(3.1)$$

Hence,
because our ultimate objective is to study
lattice measurements of physical correlation functions,
we will show in this section how to determine the behavior at large separation
$|y|$ of the pion density correlation function:
$$
\eqalignno{
     \rho(\vec{y}, t_1, t_2) &\equiv \sum_{\svec x} \langle \pi |
            \rho^u (\vec{x} , t_1)
            \rho^d (\vec{x} +\vec{y}, t_2)
            | \pi \rangle &(3.2)\cr
     &= \sum_{\svec x} \sum_n \langle \Omega_n | J^+
           (\vec{r}_{\rm out}, T)
           \,\rho^u (\vec{x}, t_1) \rho^d (\vec{x}+\vec{y},t_2) \,
          J(\vec{r}_{\rm in}, 0)  | \Omega_n \rangle
}
$$
where the up quark density operator is
$\rho^u(\vec{x},t)\equiv:~\bar{u}_{\vec{x},t}\,\gamma^0\,u_{\vec{x},t}~:$~,
$\rho^d$ is the corresponding operator for down quarks, $J(\vec{x},t)$ is the
pion source $\bar{u}_{\vec{x},t} \, \gamma_5 \, u_{\vec{x},t}$~,
$\{\Omega_n\}$ denotes states with the quantum numbers of the vacuum,
$\vec{r}_{\rm in}\/$ and $\vec{r}_{\rm out}\/$ denote the positions of the pion
sources,
and we work in Euclidean time.  The analysis of the hadron correlation function
is
analogous.

We will use the fact that at large distances, QCD is dominated by light
hadrons to motivate a tree graph approximation to the asymptotic correlation
function.  Consider a typical time history for the pion correlation function
$\rho(\vec{y})$ shown in the left portion of Figure~4, where for simplicity we
consider the equal time case $t_1 = t_2 = t$.  In the quenched approximation,
one has a $u \bar{d}$ quark-antiquark pair connecting the sources $J$,
$J^{+}$ and the density  operators $\rho^u$ and $\rho^d$
as shown by wiggly lines,
interacting via the exchange of gluons which are suppressed in the figure.
Physically, at large Euclidian separation we expect the interacting
quark-antiquark pairs to form meson ground states in the appropriate channels,
so that the quark-level time history in the left sketch is replaced by the
corresponding meson tree graph shown at the right.  As long as each meson
propagator is of sufficient length, this approximation should be accurate.
Formally, the same result may be obtained by inserting complete sets of meson
states between each of the operators and identifying the leading
singularities [10].
In either case, since the sources couple to the $\pi$ and the
external currents couple to the $\rho$, the tree-graph approximation to the
correlation function Eq.~(3.2) yields
$$
\eqalign{
             \rho_{\rm tree}({\svec y}) &=
                       \sum\limits_{\svec x}
                       \sum_{{\svec r}_1,{\svec r}_2} D_\pi
                       \left( {\svec r}_1 - {\svec r}_{\rm in} \right) D_\pi
\left( {\svec r}_2
                       - {\svec r}_1\right)
                      D_\pi \left({\svec r}_{\rm out} - {\svec r}_2\right) \cr
&\qquad\times g_{\rho\pi\pi}
\, D_\rho \left( {\svec x} - {\svec r}_1\right)
\, g_{\rho\pi\pi} D_\rho
\left( {\svec x} + {\svec y} - {\svec r}_2\right) +
\left( \hbox{``} 1\leftrightarrow2 \hbox{''}\right)
{} ~~,\cr}\eqno(3.3)
$$
where for simplicity
we use lattice propagators for scalar mesons [11] for the $D$'s.
In practice, to evaluate this expression on a lattice for comparison with
lattice QCD results, it is preferable to Fourier transform to momentum space
and perform the sums over discrete lattice momenta.

A further approximation to the sum over tree graphs in Eq.~(3.3) which
provides insight into the asymmetric behavior is given by the stationary or
classical approximation in which instead of summing over all joint positions
$r_1$ and $r_2$, we select only those joints in which the product of
propagators takes on maximal values.  Replacing the lattice propagator
by a simple continuum exponential $D(\vec{r}) \to e^{-mr}$, we obtain
$$\rho_{\rm stationary}
({\svec y}) = \sum_{\svec x}
(\rho_{\rho\pi\pi})^2
\undermax{{\svec r}_1,{\svec r}_2}
e^{-m_\pi \left| {\svec r}_1 -{\svec r}_{\rm in}\right|} \
e^{-m_\pi \left| {\svec r}_2 -{\svec r}_1\right|}\
e^{-m_\rho |{\svec x}-{\svec r}_1|}\
e^{-m_\rho\left| {\svec x}+{\svec y}-{\svec r}_2\right|} \
e^{-m_\rho \left| {\svec r}_{\rm out} -{\svec r}_2\right|} \eqno(3.4)$$
Here $|~~~|$ denotes the Euclidean distance and $r_{\rm in} = (0,{\svec 0})$,
$r_{\rm out} = (T,{\svec 0})$, and $x=(t_1,{\svec x})$).
The physical picture that results depends on the numerical value of the ratio
$m_\pi/m_\rho$.

The asymptotic behavior of pion correlation functions calculated in lattice
QCD and in the tree-level approximation are
compared in Fig.~5.  Although the lattice calculations are discussed in detail
later in Section 5, at this point the only information that is needed is the
fact that calculations were performed at three values of the hopping parameter
$\kappa$ corresponding to the hadron masses given in Table I.  For the heaviest
quark mass case, $\kappa_2$, $m_\pi$ is only 27\% below $m_\rho$ and for the
lightest case, $\kappa_5$, $m_\pi$ is slightly above ${1\over2} m_\rho$.

One important result seen in Figure 5 is the fact that beyond 1 fm, the
tree-level approximation accurately describes the behavior of the lattice QCD
results denoted by the solid curves.  The tree-level results at selected
points before image corrections are shown by the squares: Three interior
points for which image corrections are negligible, are denoted by solid
squares, and results at all other points which are subject to image effects
are denoted by open squares.  By symmetry, image contributions at the center
of the faces, center of the edges, and corners of the first unit cell may be
subtracted by dividing by 2, 4, and 8 respectively, and the results so
corrected are denoted by the solid octagons.  Thus, all the solid symbols are
free of image effects and are seen to agree extremely well with the solid
lattice QCD curves as claimed.  For the purpose of this comparison, the
image-corrected tree level results were normalized to agree with the lattice
results at $r/a = 14$.

Another significant feature observed in Fig.~5 is the fact that in the region
of 1-2 fm, the slope does not approach a fixed value given by a meson mass.
Rather, it gradually changes in this region, taking on values intermediate
between exponential decay with $m_\rho$ and $m_\pi$ indicated by the dotted
lines in the figure.  At first, this may seem surprising, since because the
current couples to the $\rho$, one might naively expect the asymptotic decay
of the correlation function at large distances to decay as $e^{-m_\rho y}$.

The behavior of the asymptotic slope may be understood qualitatively using the
stationary approximation to the tree-diagram sum.  A useful way to
state the stationary
condition for the joints $\vec{r}_1$ and $\vec{r}_2$ in Fig.~4 is to connect
all the vertices by classical strings, weight the length of each string segment
by the corresponding meson mass, and minimize the total energy of these
classical strings.  In this approximation, it is simple to determine the
asymptotic behavior in various regimes of interest.

First, consider the case $m_\pi<{1\over 2} m_\rho$, which corresponds to the
physical case and is also the simplest technically for the above model.  The
minimizing configuration is always obtained by shrinking the $\rho$-meson line
to zero and connecting sources and currents with pions, that is, the joints
coincide with the location of the currents.  Thus, for $r<\!\!<T$,
$\rho(r)\propto e^{-m_\pi r}$ while for $r>\!\!>T$, $\rho(r)\propto e^{-2m_\pi
r}$ since the pions have to go back and forth.  The case ${1\over 2}
m_\rho<m_\pi<m_\rho$ is similar except that it is energetically more favorable
for the two pions in the $t$-channel to combine into a $\rho$-meson when the
angle between them becomes smaller than the critical angle
$$
\theta_{\pi\rho} = 2\cos^{-1} \left( {m_\rho\over 2m_\pi}\right) ~~.
\eqno(3.5)
$$
Combining both cases, we conclude that one should observe a transition
from $\rho\propto e^{-m_\pi r}$ to $\rho\propto e^{-m_Hr}$ where
$m_H = \min \left\{ 2m_\pi, m_\rho \right\}$
as $r$ increases from $r<\!\!<T$ to $r>\!\!>T$.

Physically, one should note that the transition from $\rho \propto e^{-m_\pi
r}$ to $\rho \propto e^{-m_H r}$ when $r \sim T$ is a lattice artifact and
that the true physical result corresponds to the limit $T \to \infty$.  This
limit is straightforward to evaluate using the tree level path integral result
and thus provides an extremely useful extrapolation tool.  In practice, one
first calculates the tree level path integral for the actual lattice geometry
as done in Fig.~5 to verify consistency with the lattice result.  One then
recalculates the tree level path integral for $T \to \infty$, and uses the
change in the asymptotic behavior as a correction to the finite lattice
result.  The effect of this correction is shown by the dashed lines in Fig.~5
at $\kappa_2$ and $\kappa_5$, which show the correct physical asymptotic
behavior in the limit $T \to \infty$.

In principle, the tree diagram path integral analysis is analogous for the
$\rho$ and nucleon.  However, an interesting new feature arises in the case of
the $\rho$.  Naively, one would expect the relevant diagram to be of the form
shown in Fig.~4 with all three $\pi$'s replaced by $\rho$'s.
However, the operator
$\bar{u} \, \gamma_\mu \, u \simeq \omega_\mu - \rho_\mu $
does not project onto a physical $\rho$
but rather creates a linear combination of a $\rho$ and $\omega$.
The $\omega$ component may then couple to a $\rho$ and $\pi$,
since the effective hadron theory contains a $\pi \omega \rho$
vertex of the form
$$
{\cal L}_{\pi \omega \rho} = g_{\pi\omega\rho}   \partial_\alpha
\omega_\beta   \partial_\gamma  \pi   \rho_\delta   \epsilon^{\alpha \beta
\gamma \delta}
\eqno(3.6)
$$
Hence, there is also a diagram in which $\rho$'s propagate from the sources to
$\vec{r}_1$ and $\vec{r}_2$, $\omega$'s propagate from $\rho^u$ and $\rho^d$
to ${\vec r}_1$ and ${\vec r}_2$, and a $\pi$ propagates between ${\vec r}_1$
and
${\vec r}_2$.  Although
${\cal L}_{\pi\omega\rho}$ may be suppressed by the derivative couplings, at
sufficiently large $| {\vec r}_1 - {\vec r}_2 |$, the pion mass in the
propagator will
ultimately dominate the asymptotic decay, and we will observe this behavior in
the lattice results presented in Section V.

The tree level diagram model also allows one to understand the previously
puzzling discrepancy between two calculations of density-density correlation
functions in Ref.~[6].  One calculation,
which we will refer to as projected,
integrated
$\rho(\vec{x},{T\over 2}) \, \rho(\vec{x} + \vec{y}, {T\over2})$
over all $\vec{x}$
to
project onto zero momentum as in Eq.~(3.2).
The unprojected calculation used
$\rho (0,{T\over 2}) \, \rho(\vec{x}, {T\over2})$
and relied upon the fact that
non-zero momentum modes would automatically
be suppressed by evolution
for sufficiently large imaginary time $T$.

The argument is simplest in the stationary-phase approximation to the tree
level path integral, which we have verified is qualitatively similar to the
full numerical integral.
The dominant term in the sum Eq.~(3.2)
arises from ${\svec x} = -{\svec y}/2$.  If
we compare the resulting string configuration
sketched in Fig.~6
(which maximizes Eq.~(3.4)) with the
one for ${\svec x} = {\svec 0}$, one observes an important difference.
For small ${\svec y}$, the minimal configuration arises in both
cases from pure $\pi$-exchange.  Since $m_\pi > m_\rho/2$, $\rho$-mesons start
to develop in both cases for large ${\svec y}$.  However, in the unprojected
case the relevant angles are smaller and the $\rho$-meson strings
therefore
develop
earlier
as shown in the sketch.  Hence, we expect that the unprojected
correlation function will approach the $\rho$ slope earlier than the
projected one,
and this expectation is verified in the calculations shown in Fig.~7.
This result agrees qualitatively with that presented in Ref.~[6] and, we
believe, explains the reason for the observed behavior.  However, physically,
we know that the approach to the $\rho$ slope rather than the $\pi$ slope is
an artifact due to the fact that the extent in $T$ is too small.  Thus, the
most physical calculation is to perform the projection summation, which
significantly suppresses the $\rho$ contributions and then, in addition,
correct for the effect of finite $T$ by calculating the difference between the
tree diagram path integral with finite $T$ and $T \to \infty$ as discussed
in connection with the dashed curves in Fig.~5.

\goodbreak\bigskip
\hangindent=28pt\hangafter=1
\noindent{\bf IV.\quad DETERMINATION OF HADRON PROPERTIES \hfil\break
                       FROM CORRELATION FUNCTIONS ON A LATTICE}
\medskip
\nobreak
In this section we will study density-density correlations on a
lattice [12,2]
$$\rho\left( {\svec y}, t_1, t_2\right) \onequal{\rm def} \sum_{\svec x} \ll
\, h_s \,
\left| \left( \bar{u}_{{\svec x},t_2} \gamma^0 u_{{\svec x},t_2} \right)
\left( \bar{d}_{{\svec x}+{\svec y},t_1} \gamma^0 d_{{\svec x} + {\svec
y},t_1}\right) \right|
\, h_s \,
\rr\ \ ,\eqno(4.1)$$
where $|h_s\rangle$ is some superposition of states created by our lattice
sources at $t_0 = 0$ and annihilated at $t_3 = T$.  Inserting complete sets
of states one finds
$$\eqalignno{
\rho\left( {\svec y},t_1,t_2\right) &= \sum_{h_1, h_2, n} \sum_{{\svec
p},{\svec q}} e^{i{\svec q}{\svec y}} C^*_{{\svec p}, h_2} {e^{-\left( T -
t_2\right) p^0_{h_2}}\over p^0_{h_2}} &(4.2)\cr
&\qquad\times \ll h_2,{\svec p} \left| \bar{u}
\gamma^0 u \right| n,{\svec p}+ {\svec q}\rr {e^{-\left( t_2-t_1\right)
p^0_n}\over p^0_n} \ll n,{\svec p}+{\svec q}\left| \bar{d} \gamma^0 d \right|
h_1, {\svec p}\rr {e^{-t_1 p^0_{h_1}}\over p^0_{h_1}} \cdot C_{{\svec
p},h_1} \cr}$$
The kinematic factors $p^0_h = \sqrt{ {\svec p}^2 + M^2_h}$, $p^0_n = \sqrt{
\left( {\svec p}+{\svec q}\right)^2 + M^2_n}$ have been introduced to insure
that the states satisfy covariant normalization conditions in the continuum
limit.  Equation(4.2) is exact, but as it stands,
is not very useful because it is too complicated.

In addition to
summing over all excitations $n$ in the intermediate state, we have also
summed
over all excitations of the initial ($h_1$) and final ($h_2$)
states as well as over all momenta to which the lattice sources couple,
with amplitudes $C_{\svec p}$.
In this work we will use bag model sources [13] to
produce a maximal overlap with the ground state hadrons.
However, it is technically
impractical to project such localized sources to a fixed momentum such as
${\svec p} = 0$.  Since localized sources imply
an uncertainty in momentum space, one has to pay a price for
physical sources by accepting  a convolution over momenta.

In many cases, Eq.~(4.2) can be simplified considerably.  Although the bag
sources imply a convolution in momentum space, the large overlap with ground
state hadrons allows one to drop the sum over initial and final state hadrons
(provided the currents are not too close to the walls) and one is left
with the correlation function in the ground state $|h\rangle$
$$\eqalign{\rho\left( {\svec y}, t_1,t_2\right)
\approx \rho\left( {\svec y},t\right) &=
\sum_n \sum_{{\svec p},{\svec q}} {\left|C_{\svec p}\right|^2 \over p^{02}}
e^{-p_0\cdot T} \cr
&\times \ll h,{\svec p}\left| \bar{u} \gamma^0 u \right| n, {\svec p}
+ {\svec q} \rr {e^{-t\left( p^0_n-p^0\right)}\over p^0_n} \ll n, {\svec p} +
{\svec q}\left| \bar{d} \gamma^0 d\right| h,{\svec p}\rr
e^{i{\svec q}{\svec y}} ~~, \cr}
\eqno(4.3)$$
where $p_0 = p^0_h = \sqrt{ {\svec p}^2 + M^2_h}$, $t=t_2-t_1$.  As discussed
above, we are {\it a priori\/} not allowed to drop the summation over the
momentum of the initial/final state.

Let us first consider the simpler equal time case, $t=0$.
In this case,
the second moment of the density-density-correlation is
related to the size of the hadron.  For example, for
non-relativistic states one finds
$$
\ll \left( {\svec r}_u - {\svec r}_d\right)^2\rr \equiv
{\sum\limits_{\svec y} {\svec y}^2 \rho \left( {\svec y},0\right) \over
\sum\limits_{\svec y} \rho\left( {\svec y},0\right)} = - \sum\limits_n
\nabla^2_{\vec{q}^2}
F^u_{hn} \left( {\svec q}\right) F^d_{nh} \left(
- {\svec q}\right)\bigg|_{{\svec q} = 0}
{} ~~,
\eqno(4.4)
$$
where we have introduced the non-relativistic form factors
$$
F^u_{hn} \left( {\svec q}\right) = \ll h,{\svec p}\left| \bar{u}\gamma^0 u
\right| n,{\svec p} +{\svec q}\rr = \cases{
1 - {\displaystyle{R^2_u\over 6}} {\svec q}^2 + {\cal O}(\vec{q}^4) & $n=h$
\cr\noalign{\vskip 0.2cm}
{\svec q}\cdot {\svec d}^u_{hn} + {\cal O}\left( q^2\right) &$n\not=h$ \cr}
{}~~,
\eqno(4.5)
$$
and similarly for the down quarks.

Inserting the low ${\svec q}^2$ expansion of the form factors (4.5) into (4.4)
and separating the result into obvious ground state and polarization terms,
one thus obtains
$$\ll \left( {\svec r}_u - {\svec r}_d\right)^2 \rr = R^2_u + R^2_d - 2
\sum\limits_n {\svec d}^u_{hn} {\svec d}^d_{nh}
\equiv 2\left[ \ll r^2\rr_{\rm gs} + \ll r^2\rr_{\rm pol} \right] ~~.
\eqno(4.6)
$$
The origin of the dipole-dipole term,
which has been omitted previously [2,4],
becomes particularly clear if we rewrite the correlation function in the form
$$\eqalign{\ll h \left| e^{i{\svec q}\left( \hat{\svec r}_u - \hat{\svec r}_d
\right)}\right| h \rr &= \ll h \left| 1 - {\left[ {\svec q}\cdot \left(
\hat{\svec r}_u - \hat{\svec r}_d\right)\right]^2 \over 2} \right| h \rr +
{\cal O} \left( {\svec q}^4\right) \cr
&= 1-\sum\limits_n \ll h \left| {\left( {\svec q}\cdot \hat{\svec r}_u
\right)^2 \over 2} \right| n \rr \ll n|h\rr \cr
& \quad - \sum_n \ll h |n \rr\ll n \left|
{\left( {\svec q}\cdot\hat{\svec r}_d \right)^2\over 2} \right| h \rr \cr
& \quad +
\sum_n \ll h \left| \hat{\svec r}_u\cdot {\svec q}\right|n\rr \ll n \left|
\hat{\svec r}_d\cdot {\svec q}\right|h\rr + {\cal O} \left({\svec q}^4\right)
{}~~,\cr}
\eqno(4.7)
$$
where we have inserted complete sets of states in order to have only
single-particle operators appearing.  Due to the orthogonality of the
states, only $n=h$ contributes in the first two terms and these terms give
rise to the ground state rms radius in Eq.~(4.6).  The last term, however,
includes the dipole-dipole transition contributions to the correlation.

Let us
now estimate the importance of the ${\svec d}\cdot {\svec d}$ term
for nonrelativistic quarks.  From
the positivity of $\ll \left({\svec r}_u \pm {\svec r}_d\right)^2\rr$
one finds the rather crude bounds
$$- \ll r^2\rr_{\rm gs} \le \ll r^2\rr_{\rm pol} \le \ll r^2\rr_{\rm gs} \ \ .
\eqno(4.8)$$
For more realistic estimates one has to make some model assumptions.
For example, in a non-relativistic two-body problem with equal masses for up
and
down quarks one finds in the center-of-mass frame ${\svec r}_u = - {\svec r}_d$
and thus
$$\ll \left( {\svec r}_u - {\svec r}_d\right)^2 \rr = 2\left( R^2_u +
R^2_d\right)\ \ ,\eqno(4.9)$$
so that, in terms of our definitions,
$$\ll r^2\rr_{\rm pol} = \ll r^2\rr_{\rm gs} \ \ .\eqno(4.10)$$
More generally,
defining $\sum_{i<j} \langle (r_i - r_j)^2 \rangle = N (N-1)
\left[ \langle r^2 \rangle_{\rm gs} + \langle r^2 \rangle_{\rm pol} \right]$,
one finds for an $N$-body bound state with equal masses and
a symmetric coordinate space wavefunction
$$\ll r^2\rr_{\rm pol} = {1\over N-1} \ll r^2 \rr_{\rm gs} ~~.\eqno(4.11)$$
In this non-relativistic estimate, for the nucleon
$\ll r^2\rr_{\rm pol} = 1/2\ll r^2\rr_{\rm gs}$
so that neglect of the polarization correction would lead to an overestimate
of $\langle r^2 \rangle_{\rm gs}$ by 50\%
 --- certainly not a negligible correction.
The decrease of the relative importance of $\ll r^2\rr_{\rm pol}$ with the
number of constituents, Eq.~(4.11), can be understood easily.
For $N\to\infty$ the
motion of two constituents becomes more and more uncorrelated
and more accurately described by a mean field, so that
$\ll {\svec r}_i\cdot {\svec r}_j\rr \onarrow{N\to\infty}0$ for $i\not=j$.
Thus,
$\ll\left( {\svec r}_i -{\svec r}_j\right)^2\rr \to \ll r^2_i\rr
+ \ll r^2_j\rr$ for $i\not=j$.

The physical case involves significant corrections
to the previous non-relativistic argument.
A sizable fraction of the mass can come from the gluons and
the center-of-mass of the quarks need not coincide with the center-of-mass
of the hadron.  One can even imagine an extreme situation, in which the quarks
are tightly bound together and move (together) around a large cloud of gluons
--- thereby yielding a large value for $\ll r^2\rr_{\rm gs}$ while $\ll \left(
{\svec r}_i-{\svec r}_j\right)^2\rr$ remains small --- which provides a
scenario where one approaches the lower bound in Eq.~(4.8).
In addition, relativistically,
the notion of a center-of-mass is no longer
appropriate,
since the center-of-mass fluctuates.
Although we are unable to provide a realistic estimate for QCD,
the importance of the ${\svec d}\cdot{\svec
d}$-term in Eq.~(4.6) should be evident.

In a lattice calculation, one would prefer not to rely on model assumptions,
so it is helpful to consider density-density correlation functions
at unequal times [14].
Neglecting the motion of the center-of-mass for a moment, one finds in the
non-relativistic case
$$R^2(t) \equiv {\sum\limits_{\svec y} {\svec y}^2
\rho\left({\svec y},t\right)\over \sum\limits_{\svec y} \rho\left( {\svec
y},t\right)} = \ll r^2_u \rr + \ll r^2_d\rr - 2 \sum\limits_n e^{-\left( E_n -
E_h\right) t} {\svec d}^u_{hn} {\svec d}^d_{nh} ~~. \eqno(4.12)$$
Thus the ${\svec d}\cdot{\svec d}$-terms
are exponentially suppressed for large t,
which allows one not only to extract $\ll r^2_u\rr$+$\ll r^2_d\rr$ in a
model independent way but also to extract the off-diagonal
elements of the polarizability tensor
$\alpha_{ud}$(where the rows and columns are labeled
by flavor indices) from the approach to the asymptotic value by means of
$$\int^\infty_0 dt \left[ R^2(t) - R^2(\infty)\right] = - 2 \sum_n {{\svec
d}^u_{hn} {\svec d}^d_{nh}\over E_n - E_h} = 2\alpha_{ud} \ \ .\eqno(4.13)$$
A direct lattice measurement of hadron polarizabilities would be extremely
interesting in view of efforts to measure them experimentally [15].

However, as we indicated already above, there are extra complications arising
from the motion of the hadron between the two measurements as well as due to
relativistic effects.  With\footnote{*}{Here we omit the
correct relativistic normalization of ${\svec d}$ since, as we will discuss
later, we are not able to determine the polarizability in regimes where
relativistic effects become sizable.}
$$\eqalign{
&\ll \, h,{\svec p} \, \left|
\, \bar{u}\gamma^0u \, \right|
\, n,{\svec p}+{\svec q} \, \rr \cr
& \hskip.5truein = \cases{
\left(\sqrt{{\svec p}^2 + M^2_h} + \sqrt{\left( {\svec p}+{\svec q}\right)^2 +
M^2_h}\,\right) \left( 1 + {\displaystyle{R^2_u\over 6}} q^2\right) + {\cal
O}\left({\svec q}^4\right) & $n=h$ \cr\noalign{\vskip 0.2cm}
{\svec d}^u_{hn} \cdot {\svec q} + {\cal O}\left( {\svec q}^2\right)
&$n\not=h$ \cr}}
\eqno(4.14)
$$
where
$$q^2 = \left( \sqrt{{\svec p}^2 + M^2_h} - \sqrt{\left( {\svec p}+{\svec
q}\right)^2 + M^2_h}\,\right)^2 - {\svec q}^2 = - {\svec q}^2 + {\left( {\svec
p}\cdot {\svec q}\right)^2\over M^2_h + {\svec p}^2} + {\cal O} \left( {\svec
q}^3\right) ~~,
\eqno(4.15)$$
one finds for the second moment of the correlation function,
after averaging over the angular direction,
and omitting terms exponentially suppressed in $t$:
$$\eqalign{R^2_{\rm rel}(t) &\equiv {\sum\limits_{\svec y}
\rho\left( {\svec y}\right) {\svec y}^2 \over \sum\limits_{\svec y} \rho\left(
{\svec y}\right)} \cr
&= {\sum\limits_{\svec p} {\displaystyle{2C^2_{\svec p} \over E_p}}
e^{-E_p\cdot T} \left\{
2\left( R^2_u + R^2_d\right) \left[ 1 - {\displaystyle{{\svec p}^2\over
3E^2_p}}\right] + {\displaystyle{6t\over E_p}} \left[ 1 -
{\displaystyle{{\svec p}^2\over 3E^2_p}}\right]
- {\displaystyle{2t^2 {\svec p}^2\over E^2_p}}
- {\displaystyle{4{\svec p}^2\over E^4_p}} \right\} \over
\sum\limits_{\svec p} {\displaystyle{4C^2_{\svec{p}} \over E_p}} e^{-E_p\cdot
T}}  \cr} \eqno(4.16)$$
where $E_p = \sqrt{ {\svec p}^2 + M^2_h} = p^0_h$.
A comparison with Eq.~(4.12)
(non-relativistic approximation and neglect of the motion of the hadron from
$t_1$ to $t_2$) shows several new effects.  First there are some extra
relativistic normalization factors such as the factor multiplying $R^2$ which
corresponds to Lorentz contraction. Secondly, the term linear in $t$ arises
from the motion of the hadron and reflects retardation effects.  The latter
are important for the Compton-polarizability of a charged point-like
particle [15,16].  Furthermore, there are relativistic recoil terms
of order ${\svec p}^2/E^2_p$.  Most importantly, however, the
momentum ${\svec p}$ of the initial state no longer factorizes,
which makes it
cumbersome to extract rms radii, polarizabilities, and form factors
from lattice
measurements.  For example, the polarizability becomes momentum dependent.
The reason is that $\alpha_{ud}$
is defined as the second derivative of
the energy M of a particle at rest in a background electric field,
in our case
$\partial^2M/\partial {\cal E}_u \partial {\cal E}_d$,
where ${\cal E}_u$ and ${\cal E}_d$ act on up and down quarks, respectively.
Since the electric field is not a Lorentz scalar, hadrons with different
momenta ${\svec p}$ experience different polarizabilities.  The polarizability
even depends on the angle between the momentum ${\svec p}$ and the electric
field ${\svec E}$ (this dependence is contained in the ${\svec p}$-dependence
of ${\svec d}_{hn}$ which we have suppressed).

Due to these problems we will not discuss further
how to extract the polarizabilities
in the general case.
Rather, it is clearly    preferable to use momentum projected
sources --- which is a possible extension of this work
in some cases --- to extract the polarizability, as in the static case, by
integrating the
exponential term.  In practice this is again achieved by integrating the
difference between
Eq.~(4.16) and the full expression    for the   time dependent density-density
correlation function, Eq.~(4.3).

We have determined the momentum distribution $C^2_{\svec p}$
of our bag sources numerically by
calculating the overlap between a localized source and a momentum projected
source.
Typical calculations show
$ m^{-2} \langle {\svec p}^2 \rangle \equiv
m^{-2} \sum_{\svec p} C_{\svec p}^{\,2} {\svec p}^2 / \sum_{\svec p} C_{\svec
p}^{\,2} $
to be of the order of ten percent or less [17].
These are small enough that using the measured $C_{\svec p}$
one could subtract the terms linear and quadratic in $t$
and extrapolate to $t\to\infty$
in order to eliminate the exponentially suppressed dipole-dipole terms to
obtain the isoscalar rms radii
$$
{1\over2} \left( \ll r^2_u \rr + \ll r^2_d\rr \right)
= \cases{ 2\ll r^2_{\rm charge} \rr_{\pi^+}
& pion \cr\noalign{\vskip 0.2cm}
\ll r^2_{\rm charge} \rr_p + {5\over 4} \ll r^2_{\rm charge} \rr_n
& nucleon ~~.\cr}
\eqno(4.17)
$$
Note that we have restricted our attention
to $u$-$d$ correlations for the usual computational reason
that all observables can be evaluated using propagators calculated
from the bag sources.
In the future, $u$-$u$ and $d$-$d$ correlations need to be considered as well
to disentangle the electromagnetic observables in the proton and neutron [4].

\goodbreak\bigskip
\hangindent=28pt\hangafter=1
\noindent{\bf V.\quad LATTICE RESULTS}
\bigskip
\nobreak
\line{\bf Monte Carlo calculation \hfil}
\nobreak\medskip\nobreak
Density-density correlation functions were calculated on $12^3 {}\times 16$
and $16^4$ lattices using 20 quenched $SU(3)$ configurations generated by the
Cabillo-Marinari heat bath method [18] with coupling
$\beta = {6 \over g^{\,2}} = 5.7$.
For convenience, we have used the value of the lattice
spacing determined from the sting tension, $a \sim 0.2 fm$ with $a^{-1} {}\sim
1$ GeV, with the result that the spatial length of our largest lattice is 3.2
fm.  Clearly, for the purpose of comparison with the tree diagram path
integral, the precise value of the lattice scale is inessential.

Propagators were calculated at the three values of $\kappa$ shown in Table~1,
for which the lightest pion mass in approximately 340 MeV.  All the results
shown in this paper used distributed bag model sources [13,19] at the first
and last time slices, with hard-wall boundary conditions for the fermions to
prevent quark propagation across the time boundary.

The gauge fields were fixed to Coulomb gauge on the source time slices, and
the bag radius was set to 1fm.  As shown in Ref.~[13], the bag sources project
onto the hadron ground state extremely effectively, providing in general clean
signals and broad plateaus in the number of time slices away from the source.
Density operators were averaged over the central four time slices.  In order
to calculate the effect of momentum projection as mentioned in Section III,
calculations for mesons were also carried out using uniform wall sources
instead of bag sources on one time boundary.  This, of course, produced
significantly less of a plateau region on the wall source side.  Analogous
calculations for the nucleon were too noisy to be useful because integration
over the wall of a third propagator which is not tied to a density operator
requires a higher degree of phase cancellation than our present statistics
could provide.

\goodbreak\medskip
\line{\bf Image Correction \hfil}
\nobreak\smallskip\nobreak
As described in section II, the physical correlation function for an isolated
hadron is determined by subtraction of the tails of the densities of all first
images.  The most desirable way to perform this subtraction would be to
use the known asymptotic behavior of the correlator, and the tree-level path
integral described in this work provides one very attractive means to obtain
this behavior.  In analyzing the lattice results, however, we have used a
self-consistent phenomenological analysis which utilizes only the lattice data
itself.

We iteratively improve the image corrections by approximately correcting for
images using an appropriately defined parametric curve, least-squares fitting
the parameters of the curve to the corrected data, and iterating to
self-consistency.  In practice, this procedure always yields a smooth,
universal curve at large distances, as will be shown in examples below, and is
insensitive to the precise form of parameterization.  Note also that the
iterative procedure is robust and achievement of a universal curve is a strong
consistency check of the calculation and parameterization.  If the parametric
curve were too low, the image correction would be too small, and the corrected
data would then be too high and inconsistent with the curve.  The parametric
curve is thus always driven in the proper direction, and
consistency between the final curve and all the corrected data
requires correct
parameterization of the asymptotic density over a wide range of $r$.  The
actual form of the asymptotic density we have used is the following [19]
$$
\rho(r) = \rho_0 e^{-m_1 r - g(r) (r-R) (m_2-m_1)}
\eqno(5.1a)
$$
where
$$
g(r) = (1 + e^{-b (r-R)})^{-1}
\eqno(5.1b)
$$
This curve smoothly joins exponential decay with mass $m_1$ for
$r <\!\!< R$
with exponential decay with mass $m_2$ for
$r >\!\!> R$
with the transition
occurring in the vicinity of $R$ over a range $b^{-1}$.  This form has the
advantage that one can easily select physical starting values and has
sufficient flexibility for all the cases of interest in this work.

\goodbreak\medskip
\line{\bf Results \hfil}
\nobreak\smallskip\nobreak
Figure 8 shows how the self-consistent image correction procedure works in the
case of the pion density-density correlation functions.  The uncorrected data
for a $16^3$ spatial lattice are shown in (b).  By symmetry arguments, the
highest points at $r/a = 8$, $8\sqrt{2}$ and $8\sqrt{3}$
are high by factors of 2, 4, and 8 respectively,
and the other data in this regime display comparable finite
volume effects.  The self-consistent fit to the image corrected data is shown
in (c), and one clearly sees that all the data now lie on a single universal
curve which is very well fit by the self consistent solid curve specified by
Eq.~5.1.  Note that this curve is strongly constrained all the way out to the
corner of the unit cell corresponding to $8\sqrt{3}$ lattice units or $\sim
2.8 fm$.  This is the curve that was shown in Fig.~5 and agreed in detail with
the tree diagram path integral result.  For comparison, the uncorrected
lattice data for a $12^3$
\message{hey!}
spatial lattice are shown in (a), and the
self-consistent fit to the image corrected data is shown by the dashed line in
(c).  This lattice is sufficiently small that in addition to the sum of tails
of first images, one also observes errors beyond 1fm from interactions with
the images.  Given the accuracy of the tree level path integral in the region
of 1--1.5 fm, even this result on an unphysically small 2.4 fm lattice can be
combined with the tree level path integral to accurately describe the whole
correlation function.

The results for the pion density-density correlation function at all three
values of $\kappa$
given in Table I are shown in Fig.~9.  The solid curves are the
self-consistent fits to image corrected data discussed above and have
previously been shown in Fig.~5 where they agreed well with the tree level
path integral results.  For clarity in this and subsequent graphs, nearby
image-corrected lattice data are grouped into bins,
and data within each bin are
combined to a single value by means of a statistically weighted average of
both the ordinates and abscissas.  The top plot for $\kappa_2$ is thus the
binned version of plot (c) of Fig.~8.  The main conclusion from the results
of this graph is that the self-consistent fits to the image corrected data are
accurately determined for all three values of the quark mass.  In comparing
the lattice results with the slopes for the $\pi$ and $\rho$ masses, we note
that as already seen from the tree diagram results, in this regime the slope
has not yet reached the pion slope.

Analogous results for the rho density-density correlation function are shown
in Fig.~10.  In this case, we note that although one can still obtain a
meaningful self-consistent fit, the statistical errors in the lattice data are
significantly larger than for the $\pi$.  The most striking result is that as
the
quark mass decreases, the slope approaches the slope governed by the pion
mass.  We already noted in the discussion of the tree diagram analysis that
there is a $\pi \omega \rho$ effective coupling which produces a pion exchange
diagram, and subject to the limitations of the statistics, these results
indicate
that the coupling is so strong that it dominates
the decay at the separations addressed in this work.

Finally, the results for the nucleon density-density correlation function are
shown in Fig.~11.  In contrast to the meson results, the plateaus in the number
of time slices away from the wall are not always as clear as we would like,
and at present, we have no theoretical argument why this problem arises for
the nucleon.  For $\kappa_2$, the plateaus become ambiguous for
$r/a \, \grtsim \, 12$
and, indeed, one notes that the last three points appear systematically
high.  The problem is so pronounced for $\kappa_5$ that we were not
sufficiently confident of the result to present it.

The shape of the nucleon correlation function is somewhat different from that
of the mesons.  Whereas for mesons, a roughly constant asymptotic slope sets
in around 1fm, we note that only in the vicinity of 2 fm does the nucleon
slope begin to turn over and approach the rho slope.  Physically, it is
plausible that the three-quark core of the nucleon is more extended than the
quark-antiquark component of the mesons, so that when the finite size of the
relevant vertices is taken into effect, the tree diagrams only begin to
describe the asymptotic slope at substantially larger distances.

\goodbreak\bigskip
\hangindent=28pt\hangafter=1
\noindent{\bf VI.\quad SUMMARY, CONCLUSIONS, AND OUTLOOK}
\nobreak\medskip\nobreak
In this work we have analyzed and
performed measurements of
correlation functions to study the pion, the rho
and the nucleon. Our most important results are the
development and successful testing of a scheme to
correct systematically for images on periodic
lattices,
and the development of a tree level path integral to determine the asymptotic
behavior of correlation functions.
Furthermore, we have clarified the
relation between the second moment of
density-density correlators and the rms radii of hadrons.

We have shown that as long as the interaction between hadrons and
their images in adjacent unit cells is not
too large, the difference between lattice gauge
theory results for the density-density correlation function
on a periodic lattice and an infinitely large lattice
is simply due to superposition of the tails of the same correlation function
arising from mirror images.
In a similar situation in a non-gauge theory,
non-diagonal interference
terms between wave functions in adjacent cells would also contribute to  the
density-density correlations. In a gauge theory such
interference terms are proportional to Wilson loops
enclosing the periodic lattice and thus have the topology of Polyakov lines
which are strongly suppressed.
We have thus developed a practical method to use
the long distance behavior
of the density-density correlations on an infinite
lattice to correct for the images
on a finite volume lattice,
and have demonstrated its effectiveness in an
explicit lattice calculation.

QCD at large distances can be most effectively described using
hadronic degrees of freedom. As a model for the long distance
behavior we have employed tree level diagrams, including
only the lightest hadron in each channel. For the vector
current correlations measured in this work, this is equivalent
to vector meson dominance. We have used this model to
calculate the asymptotic behavior of the same correlators we have
treated on the lattice, and
observe good agreement between self-consistent fits
to the lattice data and predictions based on our tree level
hadron model.
This model also makes it
clear why the large distance behavior of density-density
correlations is not a simple exponential: since the hadron
propagates between the two vector meson vertices,
in addition to the mass scale corresponding
to
the vector
meson mass, the mass scale of the hadron itself is also relevant.
We have also shown that when the density-density correlation is measured at
spatial separations that are comparable to the temporal extent of the lattice,
there are further nontrivial r-dependencies arising from incomplete
momentum projection  due to the finite extension of the
lattice in the Euclidean time direction.
An important practical result of this work is that because
the most important finite size effects in the region $r>1.5$ fm
are now understood,
one can be confident of
future measurements of correlations
on the present physical volumes.

Nonrelativistic arguments suggest a direct relation between
the second moment of the density-density correlators
and the rms radius, as defined by the slope of the form factor.
This result arises because, nonrelativistically, the
center of mass of the quarks separates and thus the
contribution
to the density-density correlation
from dipole transitions in the intermediate state
between the two current insertions
is itself proportional to the rms radius.
Unfortunately, this argument does not apply
when the quarks are relativistic and the
relation between the second moment of the density-density
correlation and the rms-radii contains an a {\it priori}
unknown dipole-dipole term.

In this work we have only considered density-density correlations
measured at equal time.  In future work, it would be desirable to
perform measurements of density-density correlations at unequal times
for several reasons.  By measuring the
dependence of the second moment on the time difference, one could
extract the polarizability of the hadron which is physically interesting
and is being studied experimentally [15,16].
In addition,
for sufficiently large times between the two
current insertions, the dipole-dipole contributions arising from excited
intermediate states are suppressed and only the contribution from the
rms radius survives in the second moment.

\goodbreak\bigskip
\hangindent=28pt\hangafter=1
\noindent{\bf ACKNOWLEDGEMENTS}
\nobreak\medskip\nobreak
The lattice calculations were performed at the National Energy Research
Supercomputer Center under an allocation by the D.O.E.~ This work is supported
in part by funds provided by the U.S.~Department of Energy under contracts
DE-FG06-90ER40561, DE-FG06-88ER40427, and DE-FC02-94ER40818.

\vfill
\eject
\centerline{\bf APPENDIX A}
\medskip
\centerline{\bf Density-Density Correlations for}
\centerline{\bf General Currents and Dirac Particles}
\bigskip
When we derived the large $t$-behavior of the second moments of the
density-density correlation functions
we
factored the current matrix elements into the invariant form factor and a
kinematic piece $(q = p'-p)$
$$\ll p'\left|\bar{u} \gamma^0 u \right| p\rr = F_u (q^2) \cdot (p^0+p^0{}')\
\ .\eqno(\hbox{A.1})$$
Of course, for other currents one must use different kinematic factors.  For
example,
$$
\ll p'\left| \bar{u} {\svec\gamma}u\right|p\rr = F_u(q^2) \left(
{\svec p} + {\svec p}'\right)
\eqno(\hbox{A.2})
$$
or $$
\ll p'\left| \bar{u}\,u \right| p \rr = F^s_u (q^2)\ \ ,
\eqno(\hbox{A.3})
$$
where $F_u$ and $F^s_u$ are the vector and scalar up quark form factors
of the hadron.
Using these currents, the general structure of the terms appearing in the
moments of the correlation functions will be very similar to Eq.~(4.16),
although the coefficients
will be different.  For example, for $t = (t_2-t_1)$
$$\rho_{ii} \left( {\svec y},t\right) = \sum\limits_{{\svec x},i} \ll h \left|
\bar{u} \gamma^i u \left( {\svec x} + {\svec y}, t_2\right) \bar{d}\gamma^i
d\left( {\svec x},t_1\right) \right| h\rr \eqno(\hbox{A.4})$$
one finds
$$Q_{ii} \equiv \sum\limits_{\svec y} \rho_{ii} \left( {\svec
y},t\right) = \sum\limits_{\svec p} {4C^2_{\svec p} \over E^3_p} e^{-E_p\cdot
T}  {\svec p}^2 \eqno(\hbox{A.5})$$
and
$$\eqalign{
R^2_{ii} &\equiv
\sum\limits_{\svec y}
{\svec y}^2 \rho_{ii} \left( {\svec y},t\right) =
\sum\limits_{\svec p} {2C^2_{\vec p}\over E^3_p} e^{-E_p\cdot T}\cr
&\quad
\times \left\{ {2{\svec p}^2} \left( R^2_u + R^2_d\right) \left( 1 -
{{\svec p}^2\over 3E^2_p} \right)
- 3+2{\svec{p}^2\over E^2_p} \left[ 5- {3\svec{p}^2\over E^2_p}\right]
+ 2{t {\svec p}^2 \over E_p} \left[ 5 - 3 {{\svec p}^2 \over E_p^2} \right]
- 2{t^2 {\svec p}^4 \over E_p^2}
\right\} \cr}\eqno(\hbox{A.6})$$
where as usual, terms exponentially suppressed in $t$ have been omitted.
Since the $C_{\svec p}$'s, as well as the isoscalar rms Eq.~(4.17), can be
determined independently, one can actually test Eqs.~(A.5) and (A.6) by
measuring $Q_{ii}$ and $R^2_{ii}$ independently.  This provides a test of the
Lorentz invariance which we have used throughout the derivations.
Due to the coarse-grained structure of the lattice at short distances, Lorentz
invariance is not guaranteed and it will indeed break down at some point
for sufficiently large momenta.
Testing Eqs.~(A.5) and (A.6) is therefore useful to justify the Lorentz
invariance assumption made in the text for ``typical'' momenta.

We now consider the case of nucleon form factors. Here we will assume
${\svec p}=0$ in the initial state.  In practice this means that one neglects
terms of order ${\svec p}^2/M^2$ in the form factor.  Since typical nucleon
momenta
in a lattice calculation are non-relativistic, this is a reasonable
approximation.  In fact, for non-relativistic nucleons ${\svec p}^2/2M$ is of
the
order $1/T$.  Hence ${\svec p}^2/M^2$ is of the order $2/MT\approx 0.1$, where
w
e used
$M\approx 1.2\,a^{-1}$, $T=16\,a$ as typical values.

Introducing invariant form factors via Refs.~[20], [21]
$$\ll p' \left| \bar{u} \gamma^\mu u \right| p\rr = \bar{u} (p') \left[
\gamma^\mu F^u_1 (q^2)
+ {i\sigma^{\mu\nu} q_\nu\over 2M} F_2^{\,u} (q^2)\right]
u(p) \eqno(\hbox{A.7})$$
and using
$$
F_1^{\,u} (q^2) = F_1^{\,u} (0) + R_u^{\,2} {q^2 \over 6} + {\cal O}(q^4)
\eqno(\hbox{A.8a})
$$
$$
\mu_u = {1\over 2M} \left[ F^u_1(0) + F^u_2(0) \right] ~~,
\eqno(\hbox{A.8b})
$$
as well as similar expressions for down quarks, one obtains for spin-1/2
particles
$$\eqalign{{\sum\limits_{\svec y} {\svec y}^2 \rho\left({\svec y},t\right)
\over \sum\limits_{\svec y} \rho\left( {\svec y},t\right)} &=
F_1^{\,d} (0) (R_u^2 + {\textstyle 3 \over M} \mu_u)
+ F_1^{\,u} (0) (R_d^2 + {\textstyle 3 \over M} \mu_d)
+ F_1^{\,u} (0)  F_1^{\,d}(0)
\left( {\textstyle 3t \over M} - {\textstyle 3 \over 2M^2} \right) \cr
&\qquad - 2 \sum\limits_n e^{-\left( E_n - E_h\right) t}
{\svec d}^u_{hn} {\svec d}^d_{nh} ~~.\cr}
\eqno(\hbox{A.9})
$$
Note that for $\gamma^0_u$ currents the magnetic matrix elements in
Eq.~(A.7) are already of order ${\svec q}^2$, {\it i.e.\/} they do not
contribute to inelastic transitions.  Hence the polarization terms in (A.9)
are purely electric and their time integral yields the $u$-$d$ matrix element
of the electric polarizability.

The magnetic moments enter the expression for the second moment similar to the
ground state rms radii.  This is reminiscent of the Sachs form factor where
also linear combinations of $F_1$ and $F_2$ appear [21].  The term linear
in $t$ enters with the same coefficient as for spinless hadrons.  This should
not be surprising, since it arises from the classical acceleration of the
charged particle in an external electric field.  Similar to the mesonic case,
this term even survives for point-like particles. For Dirac particles
($R^2=0$, $\mu =1/2M$) the result still differs from point-like scalars and a
finite ``effective rms'' --- arising from the spin --- remains.

Of course, the non-relativistic approximation in (A.9) is not really
necessary, although, as we discussed in the context of spinless fields, the
polarizability becomes rather difficult to determine.  For the second moment,
again omitting terms exponentially suppressed in $t$,
one finds
$$
\eqalign{&{\sum\limits_{\svec y} {\svec y}^2 \rho\left( {\svec y},t\right)
\over \sum\limits_{\svec y} \rho\left( {\svec y},t\right)} = \left(
\sum\limits_{\svec p} {4{\svec C}^2_{\vec p}\over E_p} e^{-E_p\cdot
T}\right)^{-1}\cr
&\quad\times \sum\limits_{\svec p} {2{\svec C}^2_{\vec p}\over E_p}
e^{-E_p\cdot T}
\Biggl\{ 2\left( F_1^{\,d}(0) R^2_u + F_1^{\,u}(0)
R^2_d\right) \left( 1 - {{\svec p}^2\over 3E^2_p}\right)
\cr
& {\hskip1.7truein}
+ F_1^{\,u}(0) F_1^{\,d} (0)
\left[\left({3\over E^2_p}+{6t \over E_p}\right)
\left( 1 - {{\svec p}^2\over 3E^2_p}\right)- {2t^2 {\svec p}^2\over
E^2_p}\right
]
\cr
& {\hskip1.7truein}
+ {3\over E^2_p}\left[ F_1^{\,d}(0) F^u_2(0) + F_1^{\,u}(0) F^d_2(0)\right]
- {2{\svec p}^2\over E^2_pM^2} F^u_2(0) F^d_2(0)\Biggr\} \cr}
\eqno(\hbox{A.10})
$$
The only new term in (A.10) is the term quadratic in the magnetic moment ---
indicating a double spin-flip contribution to the correlation function.
\vfill
\eject
\centerline{\bf REFERENCES}

\medskip
\item{1.}K.~Barad, M.~Ogilvie and C.~Rebbi, Phys. Lett. {\bf B143}
(1984) 222; Ann.~Phys.~(NY) {\bf 168} (1986) 284.

\medskip
\item{2.}
W. Wilcox, {\it Nucl.~Phys.~B\/} (Proc.~Suppl.) {\bf 20} (1991) 459;
W.~Wilcox and K.-F. Liu, {\it Phys. Lett.\/} {\bf B172} (1986) 62;
W. Wilcox, {\it Phys. Rev.\/} {\bf D43} (1991) 2443.

\medskip
\item{3.}M.-C.~Chu, M.~Lissia and J.~W.~Negele,
{\it Nucl. Phys.\/} {\bf B360} (1991) 31.

\medskip
\item{4.}
M.-C.~Chu, M.~Lissia, J.~W.~Negele, and J.~M.~Grandy, {\it
Nucl.~Phys.~A\/} {\bf 555} (1993) 272.

\medskip
\item{5.} Lattice 93 --- Teo...
``The definition and lattice measurement of hadron wave functions,''
submitted to Proceedings of Lattice '93 (Dallas, TX)...

\medskip
\item{6.}
W. Wilcox, {\it Nucl. Phys. B\/} (Proc.~Suppl.) {\bf 26}
(1992) 491.

\item{7.}N. S. Manton, {\it Ann. Phys.\/} (NY) {\bf 159} (1985) 220; J. E.
Hetrick and Y. Hosotani, {\it Phys. Rev.\/} {\bf D38} (1988) 2621; F. Lenz, M.
Thies, S. Levit and K. Yazaki, {\it Ann. Phys.\/} (NY) {\bf 208} (1991) 1.

\medskip
\item{8.}J.~Kogut and L.~Susskind, {\it Phys. Rev.\/} {\bf D11} (1974) 395.

\medskip
\item{9.}
S.~Huang, J.~W.~Negele, and J.~Polonyi, {\it Nucl.~Phys.\/}~{\bf B307} (1988)
669.

\medskip
\item{10.}R. L. Jaffe and P. Mende, {\it Nucl. Phys. B\/} {\bf 369} (1992) 189.

\medskip
\item{11.}M. Creutz, {\it Quarks, Gluons and Lattices\/} (Cambridge University
Press, Cambridge, UK, 1983), p.~17.

\medskip
\item{12.}W. Wilcox and K.-F. Liu, {\it Phys. Rev.\/} {\bf D34} (1986) 3882.

\medskip
\item{13.}M.-C.~Chu, J.~M.~Grandy, M.~Lissia and J.~W.~Negele,
Nucl. Phys. {\bf B} (Proc. Suppl.) {\bf 26} (1992) 412.
\vfill

\medskip
\item{14.}
W.~Wilcox, {\it Phys.~Lett.\/}~{\bf B289} (1992) 411; ibid. {\it
Nucl.~Phys.~B\/} (Proc.~Suppl.) {\bf 30} (1993) 491.

\medskip
\item{15.}F. J. Federspiel {\it et al.\/}, {\it Phys. Rev. Lett.\/} {\bf 67}
(1991) 1511; J. Schmiedmayer {\it et al.\/}, {\it Phys. Rev. Lett.\/} {\bf 66}
(1991) 1015; B. R. Holstein, {\it Comm. Nucl. Part. Phys.\/} {\bf 19} (1990)
221.

\medskip
\item{16.}H. R. Fiebig, W. Wilcox and R. M. Woloshyn, {\it Nucl. Phys.\/} {\bf
B234} (1983) 47.

\medskip
\item{17.}
K.~Teo,
Ph.D. Thesis, Massachusetts Institute of Technology, 1994, unpublished.

\medskip
\item{18.}
N.~Cabibbo and E.~Marinari, {\it Phys.~Lett.\/}~{\bf B119} (1982) 387.

\medskip
\item{19.}
J.~M.~Grandy,
Ph.D. Thesis, Massachusetts Institute of Technology, 1992, unpublished.

\medskip
\item{20.}C. Ityzkson and J.-B. Zuber, {\it Quantum Field Theory\/}
(McGraw--Hill, Singapore, 1985).

\medskip
\item{21.}D. B. Leinweber, R. M. Woloshyn and T. Draper, {\it Phys. Rev.\/}
{\bf D43} (1991) 1659.

\vfill\eject

\centerline{\bf FIGURE CAPTIONS}

\medskip
\item{Fig.~1:}Density-density correlation for non-relativistic
positronium
in
QED$_{1+1}$ on a lattice with $N=10$ sites and
periodic boundary conditions.
The heavy solid curve shows the periodic solution on a circle and the
dashed curve denotes the image corrected result.
For comparison the result for
an open line $(N\to\infty)$ is shown by the light solid curve, and this
solution displaced by $\pm L$ is shown by the dotted curves to indicate
the degree of overlap.

\medskip
\item{Fig.~2:}
Fourier transforms of the density-density correlation functions shown
in Fig.~1.  The solid points denote the Fourier transform of the periodic
density at the discrete lattice momenta $p_\mu = {2 \pi k \over L}$ and the
dashed curves show the Fourier transform of the solution on the open line.
The solid curve in (a) shows the result of using the asymptotic form Eq~(2.18)
with $b/a$ chosen to make the second derivative continuous, and the solid
curve in (b) shows the errors introduced by a discontinuous second derivative
arising from Eq.~(2.16).  The solid curves in (c) are obtained as in case (a)
from an ensemble of ten data sets in which the periodic solution has been
modulated with 1\% relative random noise.

\medskip
\item{Fig.~3:}
Time histories contributing to meson density-density correlation functions
with periodic boundary conditions in the spatial direction.
The sources creating the meson are denoted by the solid dots on the
Euclidean time boundaries and the density operators are represented by
the
interior solid dots.   Quark propagators on the front and back surfaces
of the cylinder are denoted by solid and dashed wiggly lines respectively.

\medskip
\item{Fig.~4:}Space-time diagrams for $\ll\rho_u\rho_d\rr$. The left
sketch shows a typical time-history in the quenched approximation,
with quark propagators denoted by wiggly lines.  The sketch on the
right shows the corresponding tree-level graph at large separations in
which the pairs of $q\bar{q}$ propagators and the associated gluons
are replaced by ground state meson propagators.
\medskip

\item{Fig.~5:}
Comparison of approximations to the pion density-density correlation
function $\rho(r)$ at large separation $r$.  The sum of all tree-level
diagrams without image corrections is given by the open squares. Image
corrected results at the symmetry points are shown by solid octagons and
interior points with negligible image corrections are indicated by solid
squares. The best fit to the image corrected lattice QCD results is
denoted by the solid curves and for comparison, the pion and rho slopes
are shown by dotted lines.  The dashed lines indicate the corrected
asymptotic behavior obtained by taking the $T \to \infty$ limit of the
tree diagrams as discussed in the text.
The three values of the hopping
parameter $\kappa$ correspond to the masses given in Table I.  The
distance is calibrated both in lattice units (lower scale) and fm
assuming $a$=0.2fm (upper scale).

\medskip

\item{Fig.~6:}
Space-time diagrams for the pion $\ll \rho_u\rho_d\rr$ in the stationary
tree-level approximation. The upper and lower graphs correspond to small and
large spatial separations respectively. The left graphs correspond to the
configurations in which the summand in Eq.~(3.4)
is maximal and the right graphs
represent the non-summed geometry.  Note the appearance of a $\rho$
propagator in d.

\medskip
\item{Fig.~7:}Comparison between the stationary tree-level
approximation to the pion $\ll \rho_u \rho_d\rr$ in the unprojected
(full line) and the projected (dashed line) cases
for relative separation $r$.
For reference, the
slopes corresponding to
$m_\pi = 0.33$, and $m_\rho = 0.615$
are given by the dotted lines.

\medskip
\item{Fig.~8:}
Lattice results for pion density-density correlation functions with
and without image corrections.  Uncorrected results with statistical errors on
$12^3$ and $16^3$ spatial lattices are shown in (a) and (b) respectively,
shifted by a decade for clarity.  Shifted by an additional decade, the
self-consistent fit to the asymptotic density for the $16^3$ lattice using
Eq.~(5.1) is shown by the solid line in (c), along with the
image corrected lattice data
using this fit.  For comparison, the dashed
line in (c) shows the analogous self-consistent fit obtained from the $(12)^3$
data.  All results are for the hopping parameter $\kappa_2$ of table I.  The
radial distance is shown in lattice units on the bottom scale and in fm on the
top scale.

\medskip
\item{Fig.~9:}
Image corrected pion density correlation functions at the three
values of $\kappa$ given in Table I.  For clarity, the data have been
aggregated into bins, and each correlation function is shifted by 3 decades.
As in Fig.~8, the solid curves denote the self-consistent fits and the error
bars denote the image-corrected lattice measurements.  For reference, the
slopes corresponding to $m_\pi$ and $m_\rho$ given in Table~I for the
corresponding $\kappa$ are denoted by dotted lines.  The radial distance is
shown in lattice units on the bottom scale and in fm on the top scale.

\medskip
\item{Fig.~10:}
Image corrected    rho  density-density correlation    functions.   The
presentation   and notation    are the same    as in    Fig.~9.

\medskip
\item{Fig.~11:}
Image corrected nucleon density-density correlation functions.  The
presentation and notation are the same as in Fig.~9.

\par
\vfill
\eject

\centerline{\bf TABLES}

\medskip

{\narrower {\bf Table I.}~~ Hadron masses at the three values of the hopping
parameter $\kappa$ used in the lattice calculations.  For reference, a bare
quark mass $m_{\hat q} = {1\over2\kappa} - {1\over2\kappa_c}$ is also
tabulated.  The inverse lattice spacing is $a^{-1} = 1$ GeV. \bigskip}

\def\strut{\hbox{\vrule height 12pt depth 6pt width 0pt}}
\def\tstrut{\hbox{\vrule height 14pt depth 8pt width 0pt}}

\hbox to \hsize{\hss\vbox{\offinterlineskip
\hrule\halign{
\strut\vrule#&
\quad\hfil#\hfil\enskip&\vrule#&
\quad#\hfil\enskip&\vrule#&
\quad\hfil#\enskip&\vrule#&
\quad\hfil#\enskip&\vrule#&
\quad\hfil#\enskip&\vrule#&
\quad\hfil#\enskip&\vrule#\cr
& \tstrut {~}
&& \omit \hfil $\kappa$ \hfil
&& \omit \hfil ${{\displaystyle m_q}    \atop {\rm (MeV)}}$ \hfil
&& \omit \hfil ${{\displaystyle m_\pi}  \atop {\rm (MeV)}}$ \hfil
&& \omit \hfil ${{\displaystyle m_\rho} \atop {\rm (MeV)}}$ \hfil
&& \omit \hfil ${{\displaystyle m_N}    \atop {\rm (MeV)}}$ \hfil
&\cr
\noalign{\hrule}
& $\kappa_2$ && 0.16   && 175 && 691 (3) && 813 (4) && 1321 (10) &\cr
\noalign{\hrule}
& $\kappa_4$ && 0.1639 &&  95 && 511 (5) && 698 (5) && 1097 (11) &\cr
\noalign{\hrule}
& $\kappa_5$ && 0.167  &&  40 && 340 (7) && 615 (6) &&  915 (16) &\cr
}\hrule}\hss}

\bye
\end
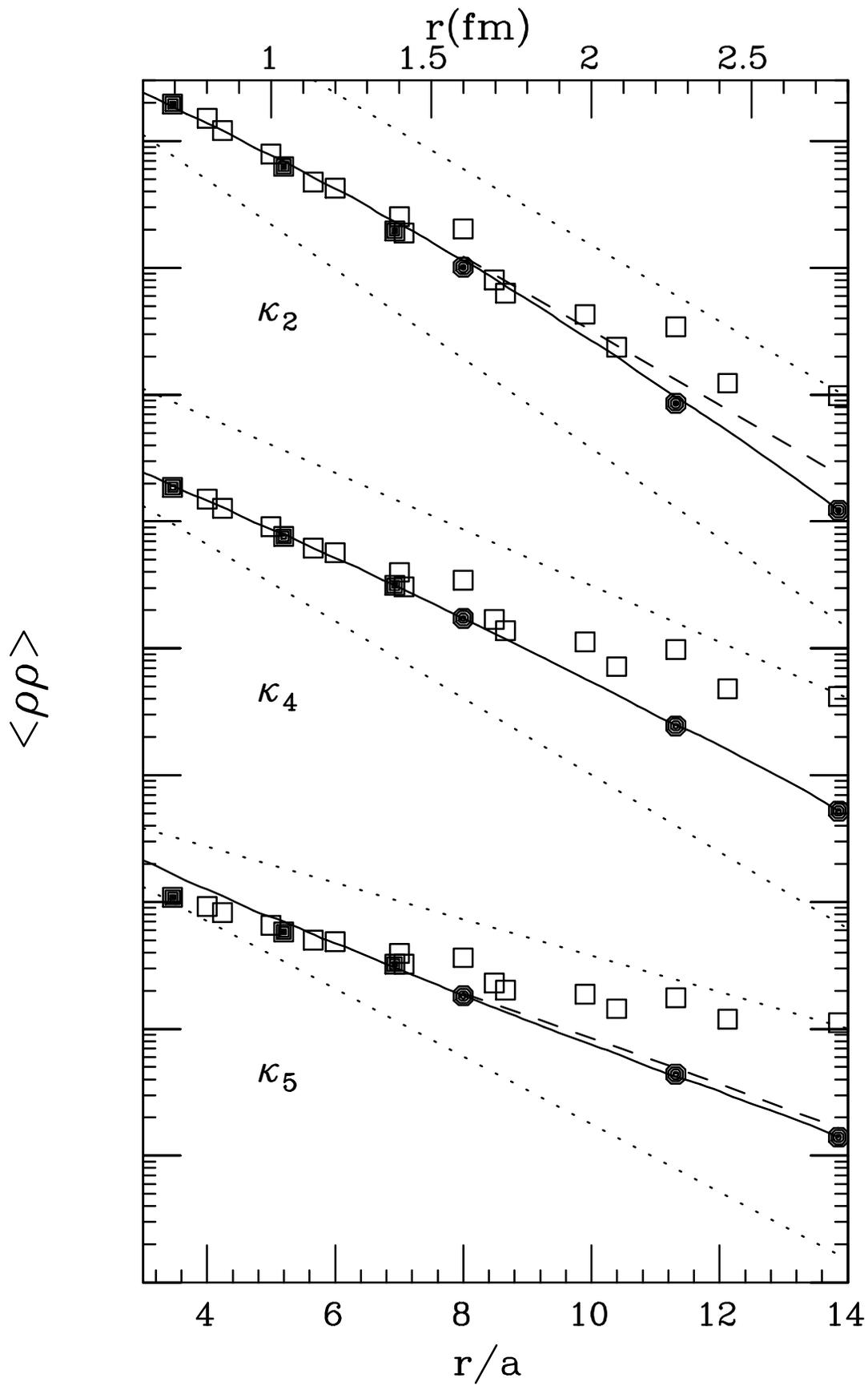
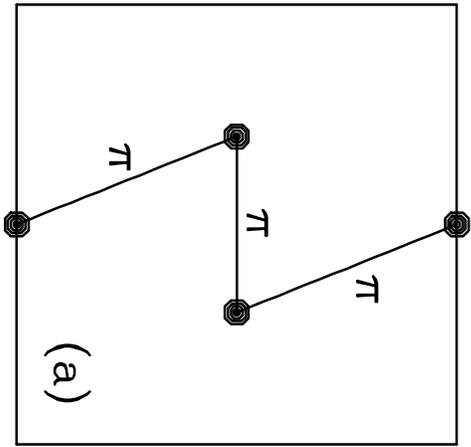
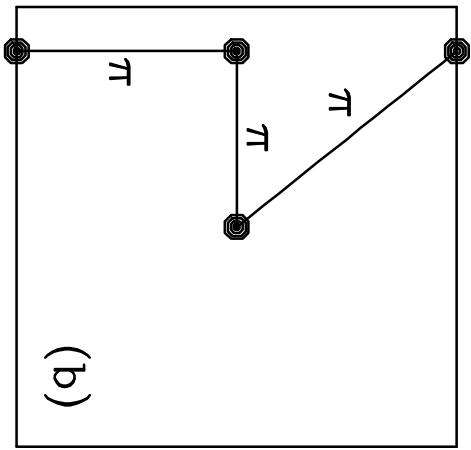
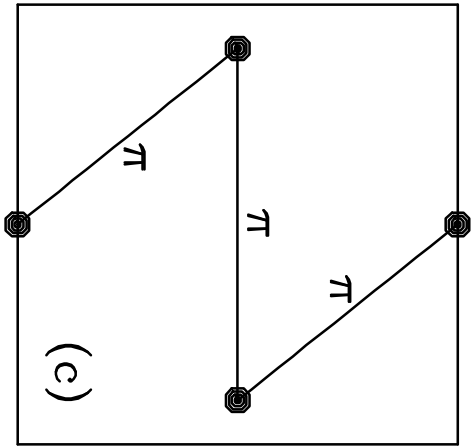
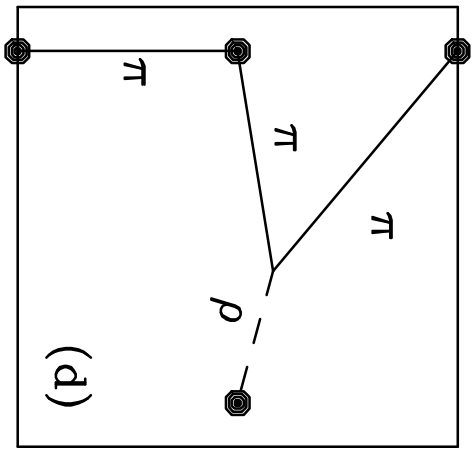
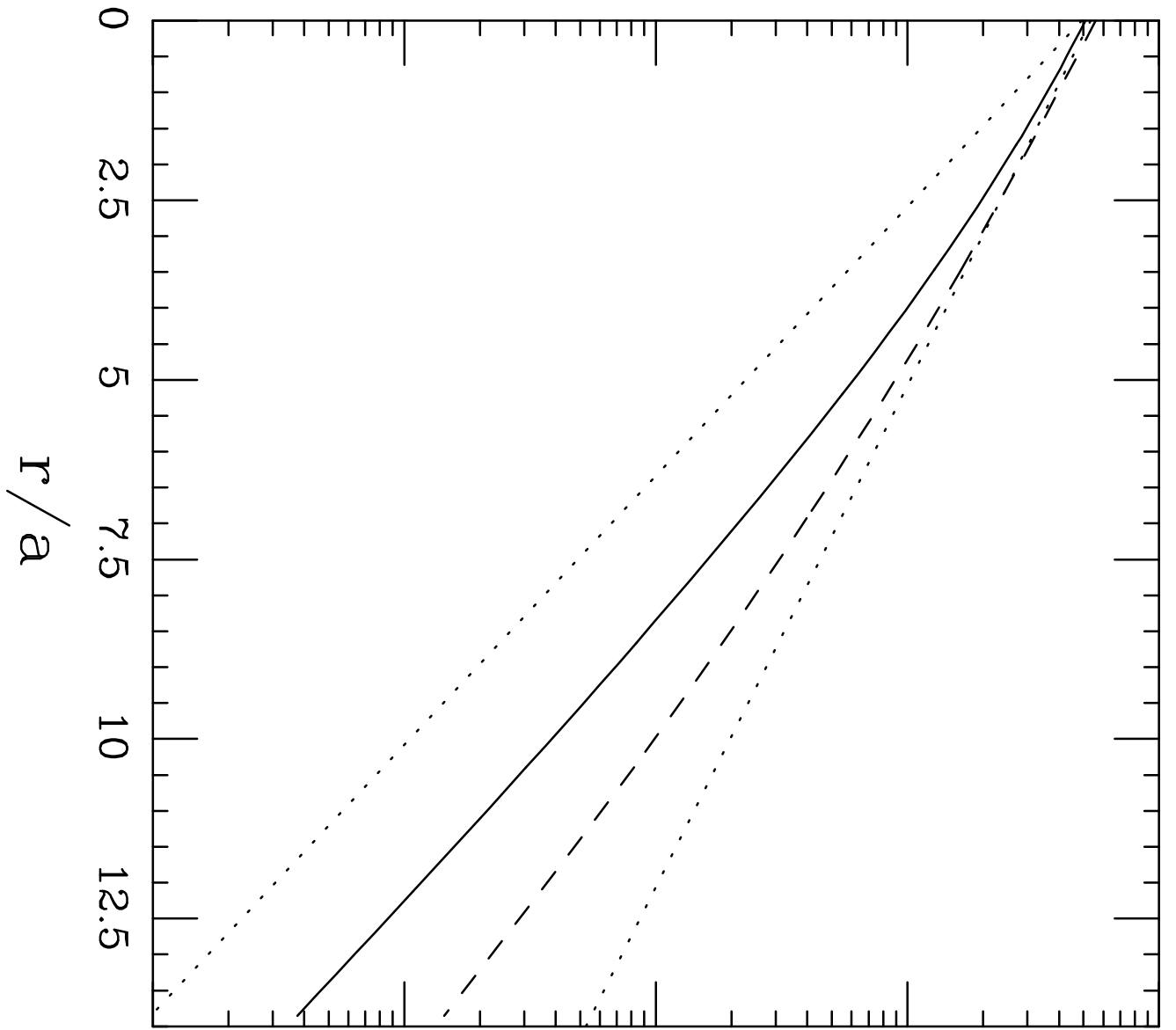
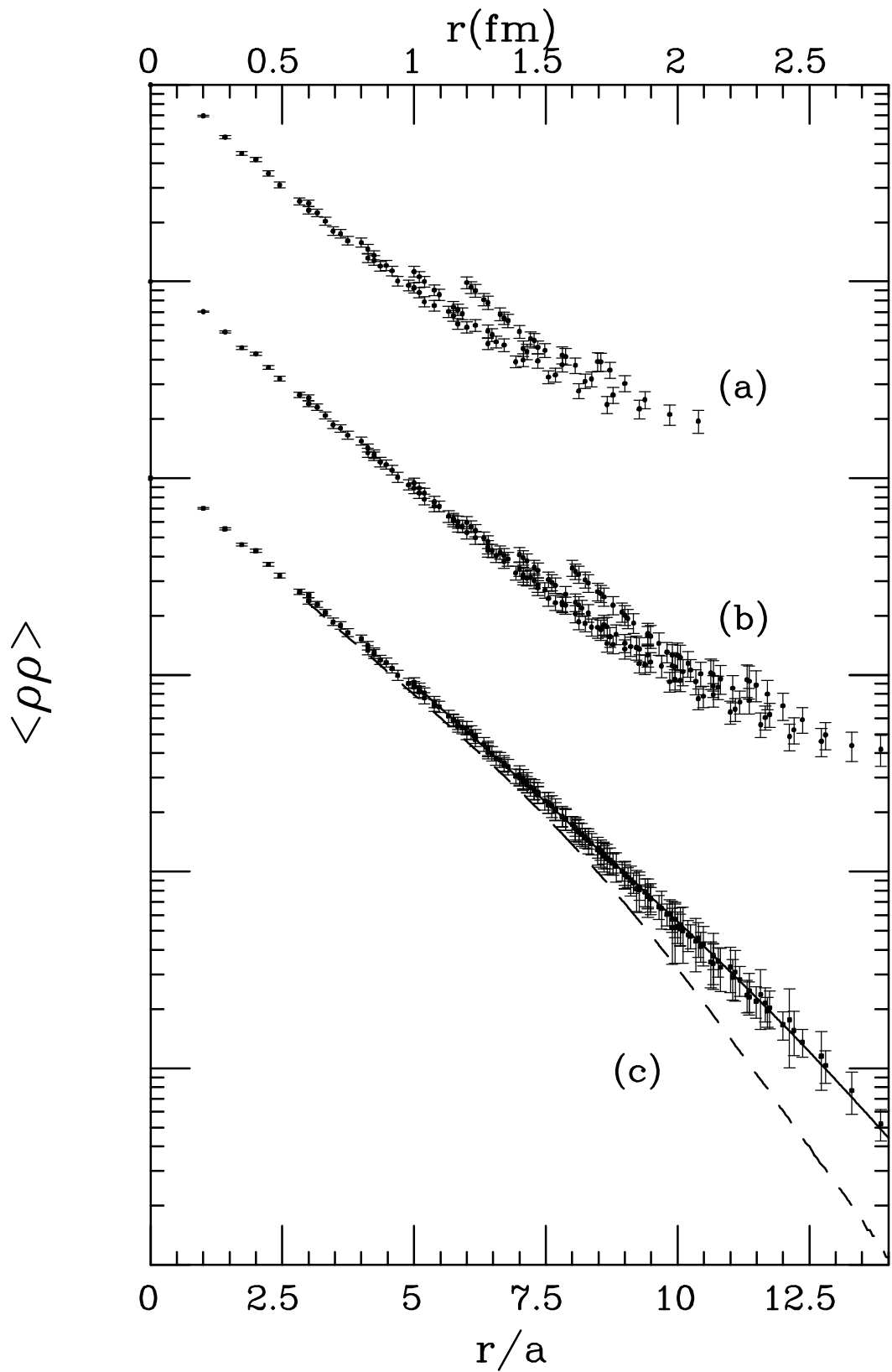
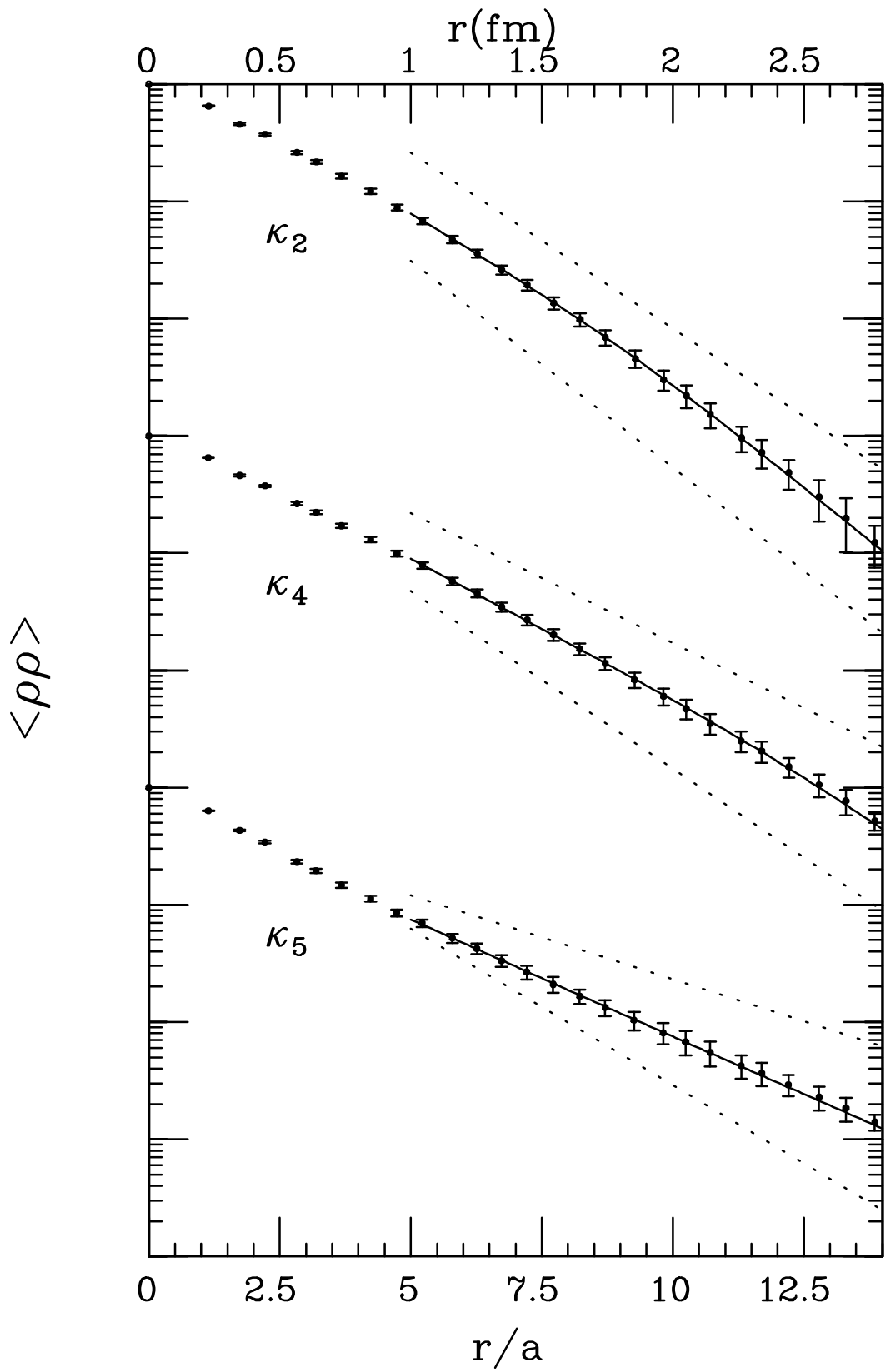
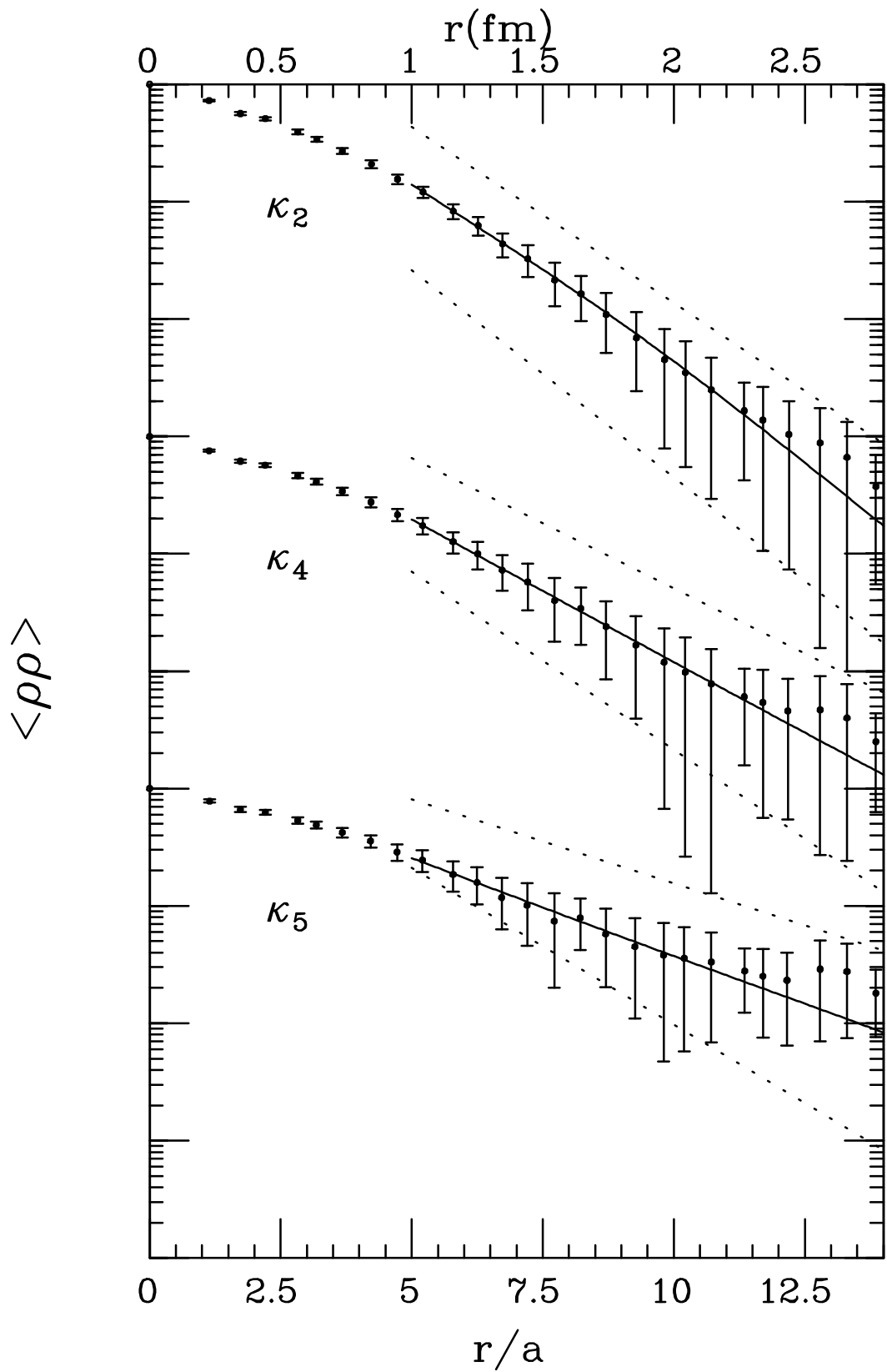
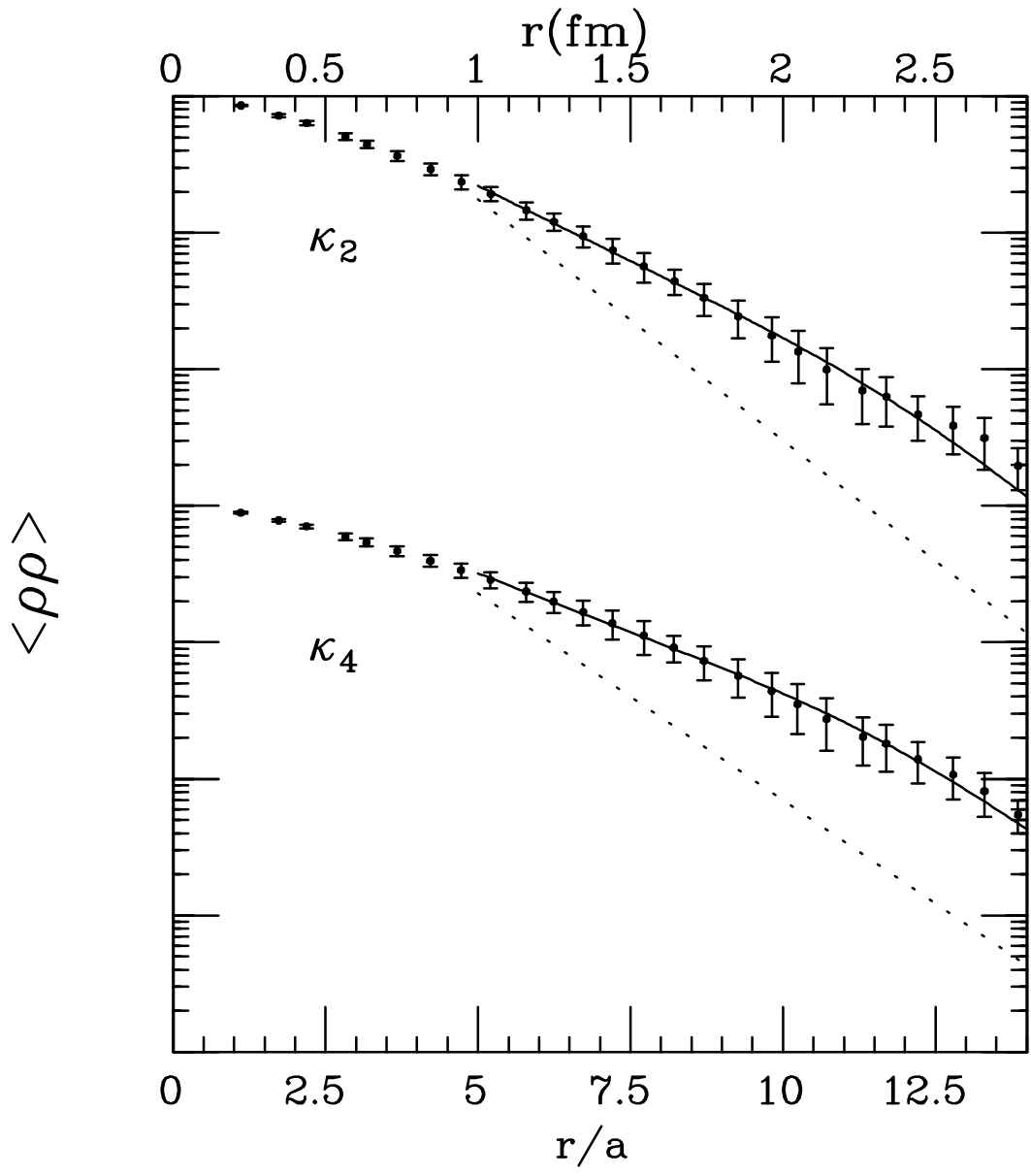
\end{document}